\documentclass[11pt,preprint]{aastex}
\usepackage{graphicx}
\usepackage{amsmath}

\def\vpsia{\vec\psi_1}

\def\da{\delta_1}
\def\db{\delta_2}
\def\l{\langle}
\def\r{\rangle}

\begin{document}
\title{A Cosmological Kinetic Theory for the Evolution of Cold Dark Matter
Halos with Substructure: Quasi-Linear Theory\\ ~ \\
}
\author{Chung--Pei Ma}
\affil{Department of Astronomy, 601 Campbell Hall,
University of California at Berkeley,
Berkeley, CA 94720; {\tt cpma@astro.berkeley.edu}}
\and
\author{Edmund Bertschinger}
\affil{Department of Physics, MIT Room 37-602A, 77 Massachusetts
  Ave., Cambridge, MA 02139; {\tt edbert@mit.edu}}

\begin{abstract}

We present a kinetic theory for the evolution of the phase-space
distribution of dark matter particles in galaxy halos in the
presence of a cosmological spectrum of fluctuations.  This theory
introduces a new way to model the formation and evolution of
halos, which traditionally have been investigated by analytic
gravitational infall models or numerical N-body methods.  Unlike
the collisionless Boltzmann equation, our kinetic equation
contains nonzero terms on the right-hand side arising from
stochastic fluctuations in the gravitational potential due to
substructures in the dark matter mass distribution. Using
statistics for constrained Gaussian random fields in standard
cosmological models, we show that our kinetic equation to
second-order in perturbation theory is of the Fokker-Planck form,
with one scattering term representing drift and the other
representing diffusion in velocity-space.  The drift is radial,
and the drift and diffusion coefficients depend only on positions
and not velocities; our relaxation process in the quasilinear
regime is therefore different from the standard two-body
relaxation.  We provide explicit expressions relating these
coefficients to the linear power spectrum of mass fluctuations and
present results for the currently favored cold dark matter model
with a nonzero cosmological constant.  Solutions to this kinetic
equation will provide a complete description of the cold dark
matter spatial and velocity distributions for the average halo
during the early phases of galaxy halo formation.

\end{abstract}

\section{Introduction}
\label{sec:intro}

Most mass in the universe is in the form of dark matter, and most dark
matter is gravitationally clustered in the form of halos.  Dark matter
halos are therefore the building blocks of the universe.  Knowledge of
the formation and evolution history of dark matter halos is essential
for an understanding of the halos' structure and dynamics, as well as
the environment that harbors galaxies and clusters.  Designs for dark
matter search experiments also rely on the predicted spatial and
velocity distributions of dark matter particles in our own galaxy.

According to the theory of structure formation in current cosmological
models, dark matter halos arise from tiny primordial fluctuations
seeded in the matter and radiation fields in the early universe at,
for example, the end of an inflationary era.  These small
inhomogeneities grow via gravitational instability, with the slightly
overdense regions becoming denser and the slightly underdense regions
becoming emptier.  At a given time, there exists a spectrum of density
fluctuations over a wide range of length and mass scales, whose
spatial distribution is characterized by the power density in Fourier
space $P(k,t)$, the power spectrum.  Dark matter halos therefore
rarely form and evolve in isolation, growing instead by frequent
accretion of smaller-mass halos and occasional major mergers with
other halos of comparable mass.

The earliest theoretical insight into cosmological halo formation
came from the spherical radial infall model, which considers an
isolated spherical overdense volume and describes the secondary
infall of bound but initially expanding collisionless matter onto
this region \citep{gg72}.  If the initial density perturbation
spectrum is scale-free and the velocity distribution follows the
Hubble law, the infall solution is found to approach a
self-similar form in an Einstein-de Sitter universe.  The
resulting density profile asymptotically approaches $\rho \propto
r^{-\alpha}$ with $\alpha\ge 2$ \citep{fg84, bert85}. The exact
value of $\alpha$ depends on the index of the initial perturbation
spectrum $P(k)\propto k^n$: it is isothermal ($\alpha=2$) for
$n\le 1 $ and for $n>1$ steepens to $\alpha\approx 3(n+3)/(n+4)$
\citep{hs85} or $3(n+3)/(n+5)$ \citep{sw98,nusser99}.  The
collapse of halos therefore retains memory of the initial
conditions in the radial infall model.

By contrast, modern cosmological N-body simulations find dark
matter halos over a wide range of mass scales to follow a density
profile that is nearly independent of the initial power spectrum
and is shallower near the halo centers (with $\alpha\sim 1$) than
that predicted by the radial infall model
\citep{nfw96,nfw97,moore98}. Furthermore, the dark matter in the
simulated halos is not entirely smoothly distributed spatially;
instead, at least 10\% of the halo mass is in the form of hundreds
to thousands of smaller, dense satellite halos of varying mass
\citep{ghigna98, klypin99, moore99}. Thus even though the analytic
similarity solutions derived from one-dimensional infall onto a
smooth overdensity have provided valuable insight into the
evolution of self-gravitating systems, realistic halo formation is
clearly more complicated.  Numerical simulation has been the tool
most frequently used in cosmology to study structure formation.
It has provided powerful insight and occasionally surprising
results (e.g. universal density profile) unexplainable by the
simulations themselves.  Our understanding of structure formation
can certainly be enhanced if a complementary tool is developed.
Complementary approaches are also especially needed at this time
to help us better understand and perhaps resolve the ``cuspy
halo'' and ``dwarf satellite'' problems of the cold dark matter
model.

In this paper we offer a new perspective on the subject of dark
matter halo formation by proposing an approach different from both
the radial infall model and N-body simulations.  We focus on the
phase-space distribution of dark matter in galaxy halos instead of
on the motion of a mass shell (as in infall models) or the orbit
of a simulation particle (as in N-body).  We attempt to understand
the evolution of the phase-space distribution with a statistical
description based on kinetic theory.  Our method is to be
contrasted with the standard method where the individual dark
matter particles are described by the collisionless Boltzmann
equation; N-body simulations are in essence a Monte Carlo method
for solving this equation. Instead, we consider the phase space of
dark matter particles in an average halo and develop a
cosmological kinetic theory for halo evolution in the presence of
a spectrum of density fluctuations.

The essential features of our approach are represented by the
fluctuation-dissipation theorem for time-dependent stochastic
processes.  To elaborate on this concept, we recall that stochastic
modelling was first developed to explain the Brownian motion of pollen
particles suspended in a fluid.  (See \S~\ref{sec:brownian} for
a more detailed discussion.)  This phenomenon is now understood to
arise from a fluctuating force on a Brownian particle due to many
rapid collisions with the molecules in the surrounding fluid.  These
fluctuations cause successive small changes in the particle velocity,
hence to a random walk.  The consequence of many such random walks is
diffusion, a form of dissipation.

Due to the large number of molecules, it is impossible to solve
the coupled equations of motion for the trajectories of all
molecules and the pollen particle in the fluid. Fortunately, a
stochastic treatment suffices.  One mathematical prescription is
provided by the Langevin equation \citep{langevin}, which models the
forces arising from the fluctuations as a stochastic term that is
added to all other deterministic forces in the equation of motion
for a particle.  An alternative prescription is given by the
Fokker-Planck equation \citep{fokker, planck}, which is an
equation of motion for the probability density or distribution
function of particles undergoing random walks.  In this case, the
collective changes in the velocities due to the fluctuations in
the system are represented by a drift coefficient and a diffusion
coefficient arising from the collision terms in the collisional
Boltzmann equation.

Stochastic phenomena occur beyond molecular scales.  In astrophysics,
the best studied example is the evolution of dense stellar systems
such as globular clusters \citep{chandra, spitzer, bt88}.  A star in a
globular cluster experiences a fluctuating force from many
gravitational two-body encounters and suffers a succession of small
velocity changes.  The fluctuations in these systems arise from
granularity in the mass distribution at the locations of the
stars. The Fokker-Planck approximation is valid for these weak
encounters. We explore this example in greater detail and contrast it
with our model of galaxy halo evolution in \S~\ref{sec:globular}.

The key feature common to all of these systems is the presence of
stochastic fluctuations that result in dissipation. In the
cosmological setting addressed by this paper, the fluctuations arise
from cosmological perturbations produced in the early universe.  At
early times, these perturbations are Gaussian and fully characterized
by their power spectrum; at late times, the substructures in halos
seen in recent simulations are the nonlinear counterparts of this
spectrum of early ripples.  These fluctuations scatter particles.  If
one considers an ensemble of halos, the masses and locations of the
substructures (or their Gaussian predecessors) are random
variables. The substructures thereby introduce ``noise'' into the
evolution of dark matter halos.  As we show in this paper, the result
is dissipation in the form of drift and velocity-space diffusion.

Recent work has explored the use of the Fokker-Planck equation in
cosmological problems.  \cite{evans97}, for example, searched for
steady-state solutions of the Fokker-Planck equation assuming the
fluctuations arise from two-body encounters.  It also assumed a simple
isotropic phase space distribution dependent only on the total energy
and a radial power law form for the potential and density. Under these
assumptions, \cite{evans97} found that $\rho\propto r^{-4/3}$ was
a steady-state solution under the effects of encounters.  There was no
dynamics in this work, however: the system was assumed to be in
collisionless equilibrium and the nature of the clumps producing the
collisions was left arbitrary.  It also remains to be determined
whether galaxy halos, either dark matter or stellar, actually evolve
to $\alpha=\frac{4}{3}$ cusp.  \cite{w00} examined the dynamics
of halos due to stochastic fluctuations from three different noise
processes for the Fokker-Planck scattering term: the merging clumps
were assumed to fly by with constant velocity, to follow decaying
orbits due to dynamical friction, or to orbit inside halos.  The
results of this study were promising and supported the notion that the
density profiles of halos were affected by mergers and substructures.
The fluctuations in this work, however, were constructed by hand
instead of being self-consistently calculated, and the results about a
central cusp were inconclusive.

Our objective here is to construct a kinetic equation in which the
drift and diffusion terms are derived from realistic density
fluctuations in current cosmological models.  We will also retain the
full phase space information by, in effect, using $f(E,J,v_r,t)$
without averaging over the angular momentum or radial velocity. We
also do not assume collisionless equilibrium; instead we base our
approach on exact dynamics.

The organization of this paper is as follows:

In \S~2, we derive the kinetic equation relevant for the evolution
of an average halo, starting with the collisionless Boltzmann equation
for individual dark matter particles.  This procedure gives rise to a
collision term on the right-hand side of the kinetic equation.  This
term represents a correlated force density and arises from
fluctuations in the ensemble averaged gravitational potential due to
clustering of the matter distribution.

In \S~3, we introduce the concepts and physical variables needed to
apply this equation to cosmological problems.  The key step is in
relating the phase space density $f(\vec r,\vec v,t)$ to the
probability densities of two standard variables in cosmology: the mass
density $\rho$ and velocity $\vec v$.  We then focus on the
quasi-linear regime in which the fluctuations about the mean density
and Hubble flow of an isotropic and homogeneous universe are
small. This focus allows us to relate $\rho$ and $\vec v$ to the
density and displacement fields $\delta$ and $\vec\psi$, which are
Gaussian random fields in standard cosmological models.

We leave the most mathematical part of the paper to
Appendix~\ref{sec:pgauss}, where we describe how to use the statistics
of constrained Gaussian random fields to express the probability
densities of \S~3 in terms of the (constrained) means and covariances
of the density and displacement fields.  The idea of a constrained
field is central to this study because our aim is to construct a
kinetic theory for dark matter particles that will later reside in a
collapsed halo rather than at any random location in the universe.
The relevant perturbations are for a density field constrained to have
certain properties, e.g., a specified height or gradient at some point
in space.  The spectrum for such constrained fields is related but not
identical to the familiar linear power spectrum $P(k)$ for
unconstrained Gaussian density fluctuations.

In \S~\ref{sec:fpeq}, we put the final expression for the correlated
force derived in Appendix~\ref{sec:pgauss} back into the kinetic
equation, and show that the resulting equation is of the Fokker-Planck
form where one term represents a drift force and the other term
represents diffusion in velocity-space. We work out these diffusion
coefficients for the currently-favored cold dark matter plus a
cosmological constant ($\Lambda$CDM) model, and provide explicit
expressions for these coefficients in terms of the linear matter
fluctuation power spectrum.  We also present the results for the
coefficients computed for this model.

\S~\ref{sec:discuss} discusses further the physical meaning of the
results in \S~\ref{sec:fpeq}.  We interpret the diffusion terms by
taking velocity moments of the kinetic equation.  To place our
cosmological kinetic equation in a broader perspective, we discuss in
more detail the classical examples of Brownian motion and globular
clusters, and point out the similarities and differences among the
three cases.  \S~\ref{sec:conclu} provides a summary and conclusions.
In Appendix~\ref{sec:stationary} we investigate the relaxation to
stationary solutions of our cosmological Fokker-Planck equation.

\section{Derivation of the Kinetic Equation}
\label{sec:kinetic}

We derive the kinetic equation for galaxy halo evolution following the
methods presented by \cite{bert93} for the derivation of the BBGKY
hierarchy.  The starting point is the one-particle phase space density
for dark matter particles in a single halo, the Klimontovich (or
Klimontovich-Dupree) density \citep{klim67,dupree}
\begin{equation}
  \label{phaseden}
  f_{\rm K}(\vec r,\vec v,t)=m\sum_i\delta_{\rm D}[\vec r-\vec r_i(t)]
    \delta_{\rm D}[\vec v-\vec v_i(t)]
\end{equation}
where $\delta_{\rm D}$ is the Dirac delta function and the
subscript K denotes the use of the Klimontovich density.  We
suppose that each halo is made of equal-mass particles of mass
$m$, for example, the elementary particles of non-baryonic dark
matter models. For convenience, we use velocity rather than
momentum and we choose units so that $fd^3v$ is a mass density. We
use proper coordinates except for \S~\ref{sec:cosmovar} and
Appendix~\ref{sec:pgauss}, where we will switch to the appropriate
cosmological (comoving) coordinates.

The Klimontovich density is simply a way of casting the exact
trajectories of all $N$ particles in a system into the form of a phase
space density.  The number of particles in $d^3r\,d^3v$ about $(\vec
r,\vec v\,)$ is $f_{\rm K}d^3r\,d^3v$, which equals either 0 or 1.
The Klimontovich density consists of delta functions because, when one
is dealing with a single system (as opposed to an ensemble), every
particle has a well-defined position and velocity.  The reason for
introducing this quantity is that it gives a particularly clean way of
going from exact N-body dynamics to a statistical description by
taking expectation values of $f_{\rm K}$ or products of more than one
$f_{\rm K}$ over an ensemble of N-body systems.  This approach to
kinetic theory is different from the usual method that begins with the
N-particle distribution for an ensemble.

The Klimontovich density obeys the Klimontovich-Dupree equation
\begin{equation}
  \label{klim}
  \frac{\partial f_{\rm K}}{\partial t}+\vec v\cdot\frac{\partial
    f_{\rm K}}{\partial\vec r}+\vec g_{\rm K}\cdot\frac{\partial
    f_{\rm K}}{\partial\vec v}=0\ ,
\end{equation}
where
\begin{equation}
  \label{gravec}
  \vec g_{\rm K}(\vec r,t)=-Gm\sum_i\frac{\vec r-\vec r_i}
    {\vert\vec r-\vec r_i\vert^3}=-G\int d^6w'\,f_{\rm K}(\vec w^
    {\,\prime},t)\frac{(\vec r-\vec r^{\,\prime})}{\vert\vec r-
    \vec r^{\,\prime}\vert^3}\ .
\end{equation}
For notational convenience we have grouped together all 6 phase space
coordinates into $\vec w\equiv\{\vec r,\vec v\}$. Eq.~(\ref{klim}) is
exact, as may be verified by substituting Eq.~(\ref{phaseden}) and
using the equations of motion for individual particles,
\begin{equation}
  \label{newton}
  \frac{d\vec r}{dt}=\vec v\ ,\quad
  \frac{d\vec v}{dt}=\vec g_{\rm K}\ .
\end{equation}
In practice, the gravitational force may be softened by modifying
Eq.~(\ref{gravec}) to reduce artificial two-body relaxation, as is
done in N-body simulations.  Eqs.~(\ref{phaseden}) and (\ref{klim})
are entirely equivalent to Eqs.~(\ref{newton}) for one system.  The
Klimontovich phase space density retains {\it all} information about a
particular halo because it specifies the trajectories of all
particles.  Eqs.~(\ref{phaseden})--(\ref{newton}) are formally
equivalent to a perfect N-body simulation.  That is not what we want.
Rather than giving a perfect description of a single halo, we average
over halos to obtain a statistical description of halo
evolution. While it is impossible to completely describe a single
galactic halo of dark matter particles, by using statistical mechanics
methods we {\it can} describe the average of infinitely many halos!

Before proceeding further, we make one important modification to the
usual procedure in kinetic theory.  Eqs.~(\ref{newton}) assume that we
work in an inertial frame.  This is satisfactory for an isolated halo
but not for a halo formed in a cosmological context where neighbors
cause it to orbit in the net gravitational field produced by all
matter. The center of mass of a halo typically moves much farther in a
Hubble time than the size of the halo itself, even in an inertial
frame in which the halo begins at rest. For example, for the cold dark
matter family of models, the displacement of a mass element from its
initial position (in comoving coordinates) is typically several Mpc
today. If we statistically average over an ensemble of halos each of
which wanders this far in a Hubble time, the phase space density will
be artificially smeared out, making the resulting average ``halo''
much larger than any realistic halo. We avoid this by doing the same
thing that simulators do when they construct halo profiles: we work in
the center-of-mass frame of each halo.

Indeed, at each moment the halo center-of-mass must be at rest in our
reference frame. This frame is non-inertial because of the
gravitational forces acting on the halo.  Thus we must modify the
kinetic equation to apply to an accelerating frame.  Fortunately, the
accelerations are small and the frame is extended only a few Mpc, so
we do not have to worry about relativistic effects.  We simply include
the inertial (``fictitious'') forces associated with working in an
accelerating frame.

We choose to define $\vec r=0$ as the center-of-mass of each halo.
Transforming Eqs.~(\ref{newton}) to the center-of-mass frame is very
simple: we replace $\vec g$ by the {\it tidal} gravitational field
relative to the center of each halo, $\vec g(\vec r\,)-\vec g(0)$, and
then subtract the center-of-mass velocity from the initial velocity of
each particle.  Eq.~(\ref{klim}) is unchanged except that $\vec g$ is
replaced by
\begin{equation}
  \label{gravt}
   \vec g_{\rm KT}(\vec r,t)=-Gm\sum_i\left(\frac{\vec r-\vec r_i}
     {\vert\vec r-\vec r_i\vert^3}+\frac{\vec r_i}{\vert\vec r_i
     \vert^3}\right)=-G\int d^6w'\,f_{\rm K}(\vec w^{\,\prime},t)
     \left(\frac{\vec r-\vec r^{\,\prime}}{\vert\vec r-\vec r^{\,\prime}
     \vert^3}+\frac{\vec r^{\,\prime}}{\vert\vec r^{\,\prime}\vert^3}
     \right)\ .
\end{equation}
The subscript T is a reminder that this is the tidal gravitational
field.

Now we return to the usual procedure for deriving a kinetic
equation. The Klimontovich density is the phase space density for a
single realization of a halo.  We might imagine performing many N-body
simulations of different halos, each with slightly different initial
conditions.  If we average the Klimontovich density over this ensemble
of halos, the result (at least, in the limit where the size of the
ensemble becomes infinite) is no longer a sum of delta functions but
instead is a smooth {\it one-particle distribution function}:
\begin{equation}
  \label{f1}
  f(\vec w,t)\equiv\left\langle f_{\rm K}(\vec w,t)\right\rangle \,,
\end{equation}
where angle brackets denote the ensemble average.  Recall that for
notational convenience we have grouped all six phase space variables
into $\vec w$.

Our goal is to derive a tractable kinetic equation for $f$.  The
obvious next step is to take the ensemble average of
Eq.~(\ref{klim}). To proceed, however, we will need the {\it
two-particle distribution function} $f_2(\vec w_1,\vec w_2,t)$, which
follows from averaging the product of Klimontovich densities at two
points:
\begin{equation}
  \label{f2}
  \left\langle f_{\rm K}(\vec w_1,t)f_{\rm K}(\vec w_2,t)
    \right\rangle\equiv\delta_{\rm D}(\vec w_1-\vec w_2)
    f(\vec w_1,t)+f_2(\vec w_1,\vec w_2,t)\ .
\end{equation}
The delta-function term $\delta_{\rm D}(\vec w_1-\vec
w_2)\equiv\delta_{\rm D}(\vec r_1-\vec r_2)\delta_{\rm D}(\vec
v_1-\vec v_2)$ corresponds to the case in which the two points lie on
top of one particle.  Two distinct particles give the $f_2$ term,
which may always be written as a product term plus a two-particle
correlation:
\begin{equation}
   \label{f2c}
   f_2(\vec w_1,\vec w_2,t)\equiv f(\vec w_1,t)f(\vec w_2,t)
     +f_{2c}(\vec w_1,\vec w_2,t)\ .
\end{equation}
This equation defines the two-point correlation function in phase
space, $f_{2c}$.

The averaging we perform is similar to the averaging used in defining
ordinary clustering correlation functions in cosmology.  The
differences here are that we are not averaging over arbitrary systems
but only over those that lead to formation of a halo at $\vec r=0$,
and that we retain velocity information.  Had we averaged over all
space and integrated over velocities, we would have gotten $\int
d^3v\,f=\bar\rho$ and $\int d^3v_1\,d^3v_2\, f_{2c}=\bar\rho^2\xi$
where $\bar\rho$ is the cosmic mean mass density and $\xi$ is the
usual spatial two-point correlation function.

Now, by taking the ensemble average of Eq.~(\ref{klim}) we arrive at a
kinetic equation for the evolution of the average halo phase space
density \citep{b96}:
\begin{equation}
  \label{boltzeq}
  \frac{\partial f}{\partial t}+\vec v\cdot\frac{\partial f}{\partial
    \vec r}+\vec g_{\rm T}\cdot\frac{\partial f}{\partial\vec v}=
    -\frac{\partial}{\partial\vec v}\cdot\vec F\ ,
\end{equation}
where we define the ensemble-averaged tidal field
\begin{equation}
  \label{g1}
  \vec g_{\rm T}(\vec r,t)\equiv\left\langle\vec g_{\rm KT}(\vec r,t)
    \right\rangle=-G\int d^6w'\,f(\vec w^{\,\prime},t)
    \left(\frac{\vec r-\vec r^{\,\prime}}{\vert\vec r-\vec r^{\,\prime}
    \vert^3}+\frac{\vec r^{\,\prime}}{\vert\vec r^{\,\prime}\vert^3}
    \right)
\end{equation}
and the correlated force density
\begin{equation}
  \label{g2}
  \vec F(\vec w,t)\equiv\hbox{Cov}\left[\vec g_{\rm KT}(\vec r,t),
    f_{\rm K}(\vec w,t)\right]=-G\int d^6w'\,f_{2c}(\vec w,
    \vec w^{\,\prime},t)\left(\frac{\vec r-\vec r^{\,\prime}}{\vert
    \vec r-\vec r^{\,\prime}\vert^3}+\frac{\vec r^{\,\prime}}{\vert
    \vec r^{\,\prime}\vert^3}\right)\ .
\end{equation}
The covariance of two random variables is denoted
$\hbox{Cov}[A,B]\equiv \l (A-\l A \r)(B-\l B \r)\r = \langle
AB\rangle-\langle A\rangle\langle B\rangle$. The product of $\vec
g_{\rm KT}$ and $f_{\rm K}$ involves a product of two Klimontovich
densities, for which we have used Eqs.~(\ref{f2}) and (\ref{f2c}).
The one-particle discreteness term in Eq.~(\ref{f2}) makes no
contribution because the particle has no self-force.

Eq.~(\ref{boltzeq}) is the first BBGKY hierarchy equation
\citep{ichimaru,dp77,peeb80}.  In cosmological applications this
equation is usually derived starting from the Liouville equation for
the N-particle distribution function.  In order to clarify the
averaging that is done here, we have started instead from the
one-particle Klimontovich density.  The result, Eq.~(\ref{boltzeq}),
is an evolution equation that looks very much like Eq.~(\ref{klim}).
By averaging over halos, however, we have introduced a correlation
integral term on the right-hand side.  The {\it average} halo
therefore does not evolve according to the Vlasov equation; instead,
it evolves according to a kinetic equation with an effective collision
term.  Eqs.~(\ref{boltzeq})--(\ref{g2}) are exact. No approximation
has been made yet.

Eq.~(\ref{boltzeq}) is a continuity equation for density in phase
space.  Without the right-hand side, it says that phase space
density is conserved along trajectories given by
Eq.~(\ref{newton}) modified for the tidal force.  The right-hand
side gives a correction due to the fact that fluctuations about
the ensemble average lead to correlated forces.  As we will see,
the effects of these fluctuations lead to dissipation.  Because
the correction term is a divergence, the mass density $\rho(\vec
r,t)=\int f\,d^3v$ is not dissipated but instead is conserved.

In most applications of kinetic theory, the right-hand side of
Eq.~(\ref{boltzeq}) arises from particle collisions. In the usual
treatment of a gas of particles with short-range forces, one assumes
$f_{2c}=0$ always except during particle collisions. This is the
assumption of molecular chaos that Boltzmann introduced to derive his
celebrated kinetic equation.  Our situation, however, is different: we
cannot neglect $f_{2c}$, because it can be much larger than the
single-particle term in Eq.~(\ref{f2c}) due to strong gravitational
clustering.

Heuristically, $f_{2c}$ describes the substructure within a galaxy
halo at the two-point level.  (Higher order correlation functions
would be needed for a complete description.)  Current cosmological
models and observations imply that the initial density field has
fluctuations that cause the formation of many small halos which
subsequently merge into present-day halos. The lumpiness of the matter
distribution represents a fluctuation about the ensemble average
density field. Through the correlated force density $\vec F$, these
fluctuations cause changes in the energy and angular momentum of
individual particle orbits that are important for the evolution of the
one-particle distribution function.

Eq.~(\ref{boltzeq}) is useful only if we find an expression for
$f_{2c}$.  One way to proceed would be to derive a kinetic
equation for $f_{2c}$ by taking the ensemble average of
Eq.~(\ref{klim}) multiplied by the Klimontovich density. This
leads to the second BBGKY equation, which depends on the
three-point correlation in phase space, which obeys the third
BBGKY equation, and so on. Thus we either have an infinite chain
of coupled equations of increasing dimensionality, or we must
close the system.  \cite{dp77} handled this problem in their
study of gravitational clustering by writing the three-point
function in terms of products of two-point functions. We will
proceed differently, by seeking to close the hierarchy at the
first level, Eq.~(\ref{boltzeq}).

To summarize this section, we have derived a formally exact
evolution equation for the average halo.  The one-particle phase
space density is conserved up to a fluctuating force term arising
from the two-point correlation function in phase space. Physically
this term represents the effects of substructure within and around
halos, which scatter particles to new orbits.  In the next two
sections, we will show that we can in fact express $f_{2c}$ in
terms of the one-particle phase space density $f$ using a relation
that is exact to second order in cosmological perturbation theory.
This relation will enable us to close the first BBGKY hierarchy.

\section{Distribution Functions from Probability Theory}
\label{sec:dfgauss}

Solving our evolution equation in Eqs.~(\ref{boltzeq})--(\ref{g2})
requires specifying the phase space density $f$ at an initial time
and the two-particle correlation $f_{2c}$ at all times.  In this
section we show how to compute these quantities from the
probability distributions of mass density and velocity at one and
two points in space.

\subsection{Probability Density vs. Phase-Space Distribution Function}

We wish to derive expressions that would relate $f(\vec w_1,t)$
and $f_{2c}(\vec w_1,w_2,t)$ to probability distributions of mass
density and velocity, because these latter quantities are
naturally specified by a cosmological model for structure
formation.  The method is to examine the mass and momentum in a
small volume of space and to construct ensemble averages of the
Klimontovich density using the probability distribution of density
and velocity.

We use the symbol $p(x,y,\ldots)$ to refer to the probability
density with respect to its arguments, where the joint probability
distribution for $(x,y,\ldots)$ is $dP(x,y,...) =
p(x,y,\ldots)\,dx\,dy\cdots$.  We assume that these probability
distributions are always normalized to unit total probability.
The variables relevant for this paper are the densities $\rho$ and
velocities $\vec{v}$ at two points $\vec{r}_1,\vec{r}_2$ and at
one time $t$: $\rho_1 \equiv \rho(\vec r_1,t)$, $\vec v_1 \equiv
\vec v(\vec r_1,t)$, $\rho_2 \equiv \rho(\vec r_2,t)$, and $\vec
v_2 \equiv \vec v(\vec r_2,t)$.  Initially we assume that the
velocity field in a given realization is single-valued in space.
We use the following probability distributions:
\begin{eqnarray}\label{probs}
   dP(\rho_1)&=&p(\rho_1)\,d\rho_1 \nonumber\\
   dP(\rho_1,\vec v_1)&=&p(\rho_1,\vec v_1)\,d\rho_1\,d^3v_1 \nonumber\\
   dP(\rho_1,\vec v_1,\rho_2)&=&p(\rho_1,\vec v_1,\rho_2)\,d\rho_1\,
     d^3v_1\,d\rho_2\nonumber\\
   dP(\rho_1,\vec v_1,\rho_2,\vec v_2)&=&p(\rho_1,\vec v_1,\rho_2,\vec v_2)\,
    d\rho_1\,d^3v_1\,d\rho_2\,d^3v_2\ ,
\end{eqnarray}
where the different probability densities are related to one another
by
\begin{eqnarray}\label{pdfs}
   p(\rho_1,\vec v_1,\rho_2)&=& \int d^3v_2\,p(\rho_1,\vec v_1,\rho_2,
    \vec v_2)     \nonumber\\
   p(\rho_1,\vec v_1)&=& \int d\rho_2\,p(\rho_1,\vec v_1,\rho_2)\nonumber\\
   p(\rho_1)&=& \int d^3v_1\,p(\rho_1,\vec v_1)\nonumber\\
    1 &=& \int d\rho_1\,p(\rho_1) \,.
\end{eqnarray}
These probability distributions depend on time but we suppress the
$t$-dependence for clarity.

For a small volume in phase space, $f(\vec w_1,t)\,d^3r_1\,d^3v_1$
equals the mean mass contained in $d^3r_1\,d^3v_1$.  The same
quantity can be calculated by averaging $\rho_1\,d^3r_1$ over
density at fixed velocity using $dP(\rho_1,\vec v_1)$, i.e., $\int
\rho_1 d^3r_1 dP(\rho_1,\vec v_1)$.  Equating the results gives
\begin{equation}\label{f1p}
  f(\vec w_1,t)=\int_0^\infty d\rho_1\,\rho_1\,
    p(\rho_1,\vec v_1)
  =\int_0^\infty d\rho_1\,\rho_1\,p(\rho_1\vert\vec v_1)\,
    p(\vec v_1)
  =\left\langle\rho_1\vert\vec v_1\right\rangle\,p(\vec v_1)\ .
\end{equation}
We have used the conditional probability density $p(\rho_1\vert\vec
v_1)$ to define the conditional mean $\langle\rho_1\vert\vec
v_1\rangle$. Note that the probability densities depend on $(\vec
r_1,t)$ through $\rho_1=\rho(\vec r_1,t)$ and $\vec v_1=\vec v(\vec
r_1,t)$, so Eq.~(\ref{f1p}) yields the desired one-particle phase
space dependence on $\vec w_1=\{\vec r_1,\vec v_1\}$.

This argument is easily extended to the two-point phase space
density, using the joint probability distribution of density and
velocity at two points.  As Eq.~(\ref{g2}) shows, we need
$f_{2c}(\vec w_1,\vec r_2,t)$, which depends on $\vec v_1$ but not
$\vec v_2$, i.e., we need the velocity at only one point.
Recalling that $f_{2c}(\vec w_1,\vec w_2,t)=f_2(\vec w_1,\vec
w_2,t) -f(\vec w_1,t)f(\vec w_2,t)$, we obtain
\begin{eqnarray}\label{f2cp}
  f_{2c}(\vec w_1,\vec r_2,t)&\equiv&\int d^3v_2\,
    f_{2c}(\vec w_1,\vec w_2,t)\nonumber\\
  &=&\int_0^\infty d\rho_1\,\rho_1\int_0^\infty d\rho_2\,\rho_2
    \left[p(\rho_1,\vec v_1,\rho_2)-p(\rho_1,\vec v_1)
    p(\rho_2)\right]\nonumber\\
  &=&\int_0^\infty d\rho_1\,\rho_1\int_0^\infty d\rho_2\,\rho_2
    \left[p(\rho_1,\rho_2\vert\vec v_1)-p(\rho_1\vert\vec v_1)
    p(\rho_2)\right]\,p(\vec v_1)\nonumber\\
  &=&\left[\left\langle\rho_1\rho_2\vert\vec v_1\right\rangle
    -\left\langle\rho_1\vert\vec v_1\right\rangle\left\langle
    \rho_2\right\rangle\right]\,p(\vec v_1)\ .
\end{eqnarray}
Although we have derived Eqs.~(\ref{f1p}) and (\ref{f2cp})
assuming a single-valued velocity field in each realization, they
are also valid when there is a distribution of velocities at each
point. One simply interprets $\rho_1$ as the total mass density
while $\vec v_1$ is a single particle velocity.  Therefore these
equations are fully general.

\subsection{Cosmological Variables}
\label{sec:cosmovar}

At high redshift, before dark matter halos formed, the density and
velocity fields were only slightly perturbed from a uniformly
expanding cosmological model.  It is thought that the fluctuations
of density and velocity responsible for all cosmic structure were
Gaussian random fields produced during the inflationary era of the
very early universe.  These fluctuations provide us with both the
initial conditions and the random element leading to a
probabilistic description for halo formation. Our goal now is to
obtain expressions for Eqs.~(\ref{f1p}) and (\ref{f2cp}) at early
times while the density fluctuations are small.

Before proceeding further, we review the description of density
and velocity fields at high redshift in an almost uniformly
expanding cosmological model.  First, the mean expansion is
described by the cosmic expansion scale factor $a(t)$ which is
related to the redshift by $1+z=1/a(t)$.  The Hubble expansion
rate is $H(t)=\dot a/a$.  The matter contributes a fraction
$\Omega_m(t)$ to the closure density. We account for the mean
expansion by defining new coordinates conventionally called
comoving coordinates: $\vec x\equiv \vec r/a(t)$.

Adopting the standard cosmological model, we assume that the dark
matter is cold.  This assumption means that the velocity field is
single-valued in space at early times before halos form.  We can
then write the density field $\rho$ and velocity field $\vec v$ in
terms of perturbations $\delta$ and $\vec\psi$ around a uniformly
expanding background:
\begin{equation}
  \label{pertdv}
  \rho(\vec r,t)=\bar\rho(t)\left[1+\delta(\vec x,t)\right]\ ,\quad
  \vec v(\vec r,t)=H(t)\left[\vec r+b(t)\vec\psi(\vec x,t)\right]\ .
\end{equation}
Here $b(t)=a(d\ln D/d\ln a)\approx a[\Omega_m(t)]^{0.6}$ is the growth
rate for density fluctuations with time-dependence $\delta(\vec
x,t)\propto D(a)$.  We emphasize that Eq.~(\ref{pertdv}) does not
assume that the density fluctuations are small; it only assumes that
the velocity field is single-valued.  We are implicitly assuming that
all of the matter may be described as a single cold non-interacting
fluid (i.e. we assume that baryons move like cold dark matter over the
length scales of interest).

The computation of the functions $a(t)$, $H(t)$, $\bar\rho(t)$,
$D(t)$, and $b(t)$ is standard in cosmology and will not be
described here. At early times when $\delta^2\ll1$, mass
conservation relates the density and velocity perturbation fields
by $\delta= -(\partial/\partial\vec x\,)\cdot\vec\psi$, or
\begin{equation}
  \label{linpsi}
  \vec\psi(\vec x,t)=-\int\frac{d^3x'}{4\pi}\,\frac{(\vec x-\vec
    x^{\,\prime})}{\vert\vec x-\vec x^{\,\prime}\vert^3}
    \,\delta(\vec x^{\,\prime},t)\ .
\end{equation}

Now we make a crucial step by changing variables from $(\vec
r,\vec v,\rho)$ used previously to $(\vec x,\vec\psi,\delta)$. The
reason for doing so is that the description in comoving
coordinates $\vec{x}$ and perturbation variables $\vec\psi$ and
$\delta$ is much simpler in the small perturbation limit.
Following the notation in \S~3.1, we write
$\delta_1\equiv\delta(\vec x_1,t)$, $\delta_2\equiv\delta(\vec
x_2,t)$, and $\vec\psi_1\equiv\vec\psi(\vec x_1,t)$.  These
variables have normalized joint probability densities
$p(\delta_1)$, $p(\delta_1,\vec\psi_1)$, and
$p(\delta_1,\vec\psi_1,\delta_2)$ in analogy with
Eqs.~(\ref{probs}) and (\ref{pdfs}).

To change variables from $(\vec v,\rho)$ to $(\vec\psi,\delta)$,
we use $d\rho=\bar\rho\,d\delta$, $d^3 v=(Hb)^3\, d^3\psi$, and
$d\rho\,d^3 v\,p(\rho,\vec v\,)=d\delta\,
d^3\psi\,p(\delta,\vec\psi\,)$.  Eq.~(\ref{f1p}) then becomes
\begin{equation}
  \label{f1p1a}
  f(\vec w_1,t)=\bar\rho\left[1+\langle\delta_1\vert\vec
    \psi_1\rangle \right] p(\vec\psi_1) \, (Hb)^{-3}\ ,
\end{equation}
where
\begin{equation}
    \langle\delta_1\vert\vec\psi_1\rangle \equiv
    \int d\delta_1\,p(\delta_1\vert\vec\psi_1)\,\delta_1\ ,\quad
    p(\delta_1\vert\vec\psi_1)=p(\delta_1,\vec\psi_1)/p(\vec\psi_1)\ ,
\end{equation}
and $p(\delta_1\vert\vec\psi_1)$ is the conditional distribution of
$\delta_1$.  Similarly, Eq.~(\ref{f2cp}) becomes
\begin{equation}
  \label{f2cp1a}
  f_{2c}(\vec w_1,\vec x_2,t)=\bar\rho^2\left[
    \langle\delta_2\vert\vec\psi_1\rangle -
    \langle\delta_2\rangle +
    \langle\delta_1\delta_2\vert\vec\psi_1\rangle -
    \langle\delta_1\vert\vec\psi_1\rangle
        \langle\delta_2\rangle
    \right]p(\vec\psi_1)(Hb)^{-3}\ .
\end{equation}

Although the derivation of Eqs.~(\ref{f1p1a}) and (\ref{f2cp1a})
assumed that the velocity field is single-valued, they are valid
in general provided that one interprets $\delta_1$ and $\delta_2$
as total mass density perturbations (not necessarily small) while
$\vec\psi_1$ is proportional to the single particle velocity.
Eqs.~(\ref{f1p1a}) and (\ref{f2cp1a}) are equivalent to
Eqs.~(\ref{f1p}) and (\ref{f2cp}) expressed in different
variables. Their utility becomes evident in
Appendix~\ref{sec:pgauss} where we evaluate the probability
distributions of density and velocity in cosmological models.

All that remains is to calculate the expectation values in
Eqs.~(\ref{f1p1a}) and (\ref{f2cp1a}) and to substitute the latter
into Eq.~(\ref{g2}).  This calculation is somewhat complicated and
is presented in Appendix~\ref{sec:pgauss}.  Part of the
complication arises because of an important detail we have so far
neglected: halos form preferentially in regions that are initially
overdense. Our calculations assume that a halo forms at $\vec
r=0$. Consequently we should not evaluate the expectation values
for arbitrary points in space but rather for those that satisfy
appropriate constraints.

Strictly speaking it is impossible to specify in advance the regions
that will form halos in a cosmological model with random initial
fluctuations because the dynamics of gravitational clustering is too
complicated.  A simple model, however, has been found to give
reasonable agreement with numerical simulations: halos form at maxima
of the smoothed initial density field \citep{bbks}.
These constraints are found to modify the expectation values in
Eqs.~(\ref{f1p1a}) and (\ref{f2cp1a}).  To emphasize the application
of constraints, we rewrite these expectation values with a subscript
$C$:
\begin{equation}
  \label{f1p1}
  f(\vec w_1,t)=\bar\rho\left[1+\langle\delta_1\vert\vec
    \psi_1\rangle_C \right] p(\vec\psi_1) \, (Hb)^{-3}\ ,
\end{equation}
and
\begin{equation}
  \label{f2cp1}
  f_{2c}(\vec w_1,\vec x_2,t)=\bar\rho^2\left[
    \langle\delta_2\vert\vec\psi_1\rangle_C -
    \langle\delta_2\rangle_C +
    \langle\delta_1\delta_2\vert\vec\psi_1\rangle_C -
    \langle\delta_1\vert\vec\psi_1\rangle_C
        \langle\delta_2\rangle_C
    \right]p(\vec\psi_1)(Hb)^{-3}\ .
\end{equation}
Eqs.~(\ref{f1p1}) and (\ref{f2cp1}) are the main results of this
section.  We will use them to specify the initial conditions on $f$
and to determine the correlated force density in Eq.~(\ref{g2}).

The next step is to evaluate the necessary constrained expectation
values using the standard cosmological model.  We present the details
of this calculation in Appendix~\ref{sec:pgauss}, where we choose to
err on the side of simplicity by following \cite{bbks} and using only
the zeroth and first derivatives of the smoothed initial density field
$\Delta(\vec x\,)$ as constraints.  Detailed comparison has shown that
this simple model does not predict perfectly all halo formation sites
\citep{kqg}, and more complicated constraints work better,
e.g. \cite{monaco}.  The procedure of Appendix~\ref{sec:pgauss} can
presumably be extended to incorporate more complicated constraints.

\section{Fokker-Planck Equation}
\label{sec:fpeq}

After having devoted \S~\ref{sec:dfgauss} and
Appendix~\ref{sec:pgauss} to probability and statistics we are now
ready to return to the kinetic equation in \S~\ref{sec:kinetic}
and write down an explicit expression for the correlated force
density $\vec F(\vec w,t)$ on the right-hand-side of the kinetic
equation (\ref{boltzeq}).  The approach we have followed is exact
rather than phenomenological --- our starting point was the first
BBGKY hierarchy equation, an exact kinetic equation.  The only
approximation we have made is to assume that the density and
velocity fields are Gaussian random fields.  The resulting
two-particle correlation function, Eq.~(\ref{f2cexact}), depends
in a simple way on the one-particle distribution $p(\vec\psi_1)$.
As we will show next, this simple dependence leads to the
Fokker-Planck equation.  This is a remarkable and non-trivial
result.  The Fokker-Planck equation is a great simplification of
the exact --- and intractable --- BBGKY hierarchy.  Usually the
Fokker-Planck equation is {\it assumed} to hold a priori, and then
its diffusion coefficients are estimated.  Here, on the other
hand, we have {\it derived} the Fokker-Planck equation.  Along
with it we obtain the diffusion coefficients in the quasi-linear
regime, as follows.

\subsection{Diffusion Coefficients}
\label{sec:fpdiffco}

Substituting Eq.~(\ref{f2cexact}) into Eq.~(\ref{g2}) using
Eq.~(\ref{f2cp}), converting to comoving coordinates, using
Eq.~(\ref{linpsi}) to relate $\vec\psi$ to $\vec v$, and using
Eq.~(\ref{f1lin}), to third order in $\delta$ we get
\begin{eqnarray}
  \label{g2lin}
  \vec F(\vec w,t) &=& -4\pi G\bar\rho a \left\{ \left[
    C(\delta,\vec\psi_0) - C(\delta,\vec\psi\,)\cdot
    C^{-1}(\vec\psi,\vec\psi\,)\cdot C(\vec\psi,\vec\psi_0)\right] \,f \right.
      \nonumber\\
     &&  + \left. (Hb) \left[ C(\vec\psi,\vec\psi\,)-C(\vec\psi_0,\vec\psi\,)
       -\langle\vec\psi\,\rangle_C\otimes C(\delta,\vec\psi\,)
    \right] \cdot\frac{\partial f}{\partial\vec v}\right\}+O(\delta^4)\ ,
\end{eqnarray}
where the covariance matrices $C$ of variables $\delta$ and
$\vec\psi$ are given in Eqs.~(\ref{covfun}) and (\ref{c0}).  Here,
$\vec\psi\equiv\vec\psi(\vec x,t)$ and
$\vec\psi_0\equiv\vec\psi(0,t)$.  Note that all the terms with
subscript 0 arise from our use of the tidal field rather than the
total gravitational acceleration.

Substituting Eq.~(\ref{g2lin}) into Eq.~(\ref{boltzeq}), in proper
coordinates our kinetic equation in the quasilinear regime reduces
to the Fokker-Planck equation
\begin{equation}
  \label{fplin}
  \frac{\partial f}{\partial t}+\vec v\cdot\frac{\partial f}{\partial
    \vec r}+\vec g_{\rm T}\cdot\frac{\partial f}{\partial\vec v}
  =-\frac{\partial}{\partial\vec v}\cdot\left(\vec A\,f-{\bf D}
    \cdot\frac{\partial f}{\partial\vec v}\right)
\end{equation}
specified by a vector field $\vec A$ and a tensor field ${\bf D}$.
Together, $\vec A$ and ${\bf D}$ are called diffusion
coefficients. Heuristically, $\vec A$ represents a drift force ($\vec
A$ has units of acceleration) while ${\bf D}$ represents the diffusive
effects of fluctuating forces.  The flux density in velocity space is
$\vec Af-{\bf D}\cdot\partial f/\partial\vec v$.  Keeping only terms
up to second order in the density perturbations, these quantities are
given by
\begin{eqnarray}
  \label{driftlin}
  \vec A(\vec r,t)&\approx&-4\pi G\bar\rho a\left[
    C(\delta,\vec\psi_0)-C(\delta,\vec\psi\,)\cdot C^{-1}
    (\vec\psi,\vec\psi)\cdot C(\vec\psi,\vec\psi_0\,)\right]
    \nonumber\\
  &=&- \hbox{Cov}(\delta,\vec g_0\vert\vec\psi\,)
    =  \hbox{Cov}(\delta,\vec g_{\rm T}\vert\vec v\,)
\end{eqnarray}
and
\begin{equation}
  \label{difflin}
  {\bf D}(\vec r,t)\approx4\pi G\bar\rho a Hb
    C(\vec\psi-\vec\psi_0,\vec\psi\,)
    = \hbox{Cov}(\vec g_{\rm T},\vec v\,)\ .
\end{equation}
In Eqs.~(\ref{driftlin}) and (\ref{difflin}) we have also used the
fact that in the quasilinear regime the peculiar gravity and
displacement are related by $\vec g=4\pi G\bar\rho a\,\vec\psi$
\citep{zel70}, and the tidal field is $\vec g_{\rm T}\equiv\vec
g(\vec r,t)-\vec g_0$, where $\vec g_0\equiv\vec g(0,t)$.  In
writing the second line of Eq.~(\ref{driftlin}) we have used
Eq.~(\ref{covy}) with $Y_B=\{\delta,\vec\psi_0\}$ and
$Y_A=\vec\psi-\l\vec\psi\r_C$. Thus, $\hbox{Cov}(\delta,\vec
g_0\vert\vec v\,)$ should be understood to be subject to the same
extremum or other constraints as e.g. the constrained covariance
function $C(\delta,\vec\psi\,)$.  The displacement and velocity
are related by the second of Eqs.~(\ref{pertdv}).  Thus, in linear
theory $\vec g\propto\vec v$ so that conditioning on $\vec v$
makes the fluctuation in $\vec g$ vanish, allowing us to replace
$- \hbox{Cov}(\delta,\vec g_0\vert\vec\psi\,)$ by
$\hbox{Cov}(\delta,\vec g_{\rm T}\vert\vec v\,)$ in
Eq.~(\ref{driftlin}).  In all cases, the covariances are taken
subject to the extremum constraints discussed in
Appendix~\ref{sec:pgauss}.

It is worth noting that had we used the total gravitational
acceleration $\vec g(\vec r,t)$ instead of the tidal field $\vec
g_{\rm T}(\vec r,t)$, the drift vector $\vec A$ would have been
zero and ${\bf D}=\hbox{Cov}(\vec g,\vec v\,)$ would have been
larger. We conclude that tidal effects are crucial for a proper
description of halo evolution \citep{dekel03}.  This is evident in
the final expressions for $\vec A$ and ${\bf D}$.  After the
lengthy derivations, these expressions are remarkably simple,
suggesting that they may have greater validity than the case of
small-amplitude Gaussian fluctuations that we have analyzed here.

In going from Eqs.~(\ref{f1lin}) and (\ref{g2lin}) to
Eqs.~(\ref{driftlin}) and (\ref{difflin}) we have dropped the
contributions made by $\langle\delta\rangle_C$ and $\langle
\vec\psi\,\rangle_C$ because they do not contribute at second
order in $\delta$.  At this order, $\vec A$ and ${\bf D}$ are
independent of the value of the smoothed density $\Delta_0$ used
for the constraint. However, the constraint enters in third order.

Carrying out the derivation to one higher order in $\delta$
requires using the following result for Gaussian random variables:
$\hbox{Cov}(AB,C)=\langle A\rangle\hbox{Cov}(B,C)+ \langle
B\rangle \hbox{Cov}(A,C)$.  Applying this with
$B=(1+\delta)^{-1}=1-\delta+O(\delta^2)$, after some algebra we
find
\begin{equation}\label{diffuns}
  \vec A(\vec r,t)=\hbox{Cov}\left(\delta,\frac{\vec g_{\rm T}}
    {1+\delta}\,\Big\vert\,\vec v\,\right)+O(\delta^4)\ ,\quad
  {\bf D}(\vec r,t)=\hbox{Cov}\left(\frac{\vec g_{\rm T}}
    {1+\delta},\vec v\,\right)+O(\delta^4)\ .
\end{equation}
Although these results are correct for Gaussian fluctuations to
third order in the density perturbations, nonlinear evolution will
generate non-vanishing irreducible three-point correlations (i.e.,
non-Gaussianity) that have been neglected here.  For example,
nonlinear evolution generates corrections to the Zel'dovich
approximation so that velocity and gravity are no longer
proportional to each other as we have assumed.  For this reason,
Eqs.~(\ref{diffuns}) should not be regarded as being more accurate
than Eqs.~(\ref{driftlin}) and (\ref{difflin}).

Despite their limitations, Eqs.~(\ref{diffuns}) are useful in
showing the qualitative effect of applying an initial constraint.
Choosing the proto-halo to have $\Delta_0>0$ means that typically
$\delta>0$ in the proto-halo, which decreases $\vec A$ and ${\bf
D}$ compared with Eqs.~(\ref{driftlin}) and (\ref{difflin}).
Conversely, the diffusion coefficients are enhanced in void
regions.  Note also that the constraint value appears explicitly
in the initial conditions for $f$ in the linear regime through the
term $\l\delta\r_C$ in Eq.~(\ref{f1lin}).

Eqs.~(\ref{driftlin}) and (\ref{difflin}) imply that, in the
quasilinear regime, the drift and diffusivity are independent of
the velocity within the halo.  (Although $\vec A$ is equivalent to
a covariance at fixed $\vec v$, it is independent of the
particular value of $\vec v$.)  This follows because $f_{2c}$
depends on $\vec\psi_1=(\vec v_1-H\vec x_1\,)/(Hb)$ only through
$p(\vec\psi_1)$ and $\partial p/\partial\vec \psi_1$ in
Eq.~(\ref{f2cexact}).  As noted above, at second order in
perturbation theory the drift and diffusivity are also independent
of the initial overdensity of the peak $\Delta_0$ (but they depend
on the smoothing scale used to find extrema). Thus, both the
Fokker-Planck equation and the diffusion coefficients appear to be
remarkably robust quantities.

To make further use of Eqs.~(\ref{driftlin}) and (\ref{difflin}),
we need to write out the coefficients explicitly.  Using
Eqs.~(\ref{covfun}), (\ref{driftlin}), and (\ref{difflin}) we
first factor out the dependence on cosmology by writing
\begin{subequations}\label{adscaled}
\begin{eqnarray}
   \vec{A}(\vec r,t) &=& 4\pi G\bar\rho a\,A_r(r,t)\hat r\,,\\
   {\bf D}(\vec r,t) &=& 4\pi G \bar\rho a H b\, [D_r(r,t)\hat
    r\otimes\hat r+D_t(r,t)({\bf I}-\hat r\otimes\hat r)] \,,
\end{eqnarray}
\end{subequations}
where $D_r$ and $D_t$ are scaled radial and tangential diffusion
coefficients.  The scaled variables are
\begin{subequations}\label{coeffs}
\begin{eqnarray}
  A_r(r,t) &=& -r\eta(r) - \frac{\bar\eta(0)}{\sigma_1^2}\frac{d\bar\xi(r)}
    {dr}+\frac{1}{c_r(r)}\frac{d\gamma(r)}{dr} \left\{\frac{r\bar\xi(r)
    \bar\eta(r)}{\sigma_0^2}+\frac{1}{\sigma_1^2}\frac{d\bar\xi(r)}{dr}
    \left[\bar\xi(r)-2\bar\eta(r)\right]\right\}  \ , \label{scala}\\
   D_r(r,t) &=& c_r(r) - \frac{d\gamma(r)}{dr}=
     \sigma_\psi^2-\frac{d\gamma(r)}{dr}-\left[\frac{r\bar\eta(r)}
     {\sigma_0}\right]^2-\left[\frac{\bar\xi(r)-2\bar\eta(r)}{\sigma_1}
     \right]^2\ , \label{scaldr}\\
   D_t(r,t) &=& c_t(r) - \frac{\gamma(r)}{r}=
     \sigma_\psi^2-\frac{\gamma(r)}{r}-\left[\frac{\bar\eta(r)}{\sigma_1}
     \right]^2 \ . \label{scaldt}
\end{eqnarray}
\end{subequations}
Here, $r$ is the comoving distance from the center of the halo.
The time dependence arises only because the functions appearing in
Eqs.~(\ref{coeffs}) depend on the time-dependent power spectrum of
density fluctuations.  These functions are summarized here for
convenience:
\begin{subequations}\label{sumfun}
\begin{eqnarray}
  \eta(r) = \int\frac{d^3k}{(2\pi)^3}\,P(k)\frac{j_1(kr)}{kr}\ ,
    \ \
    \gamma(r) = \int\frac{d^3k}{(2\pi)^3}\,P(k)\frac{j_1(kr)}{k^3}\ ,
    ~~~~~~~~~~~~~~~~~~~~~~~~~~~~~~~~~~~~~~~
      &&\label{etagamma}\\
  \bar\xi(r) = \int\frac{d^3k}{(2\pi)^3}\,P(k)W_R(k) j_0(kr)\ ,
    \ \
    \bar\eta(r) = \int\frac{d^3k}{(2\pi)^3}\,P(k)W_R(k)
      \frac{j_1(kr)}{kr}\ ,~~~~~~~~~~~~~~~~~~~~~~~
      &&\label{xibareta}\\
  \sigma_0^2 = \int\frac{d^3k}{(2\pi)^3}\,P(k)W_R^2(k)\,,
  \ \
  \sigma_1^2 = \frac{1}{3}\int\frac{d^3k}{(2\pi)^3}\,k^2 P(k)W_R^2(k)\,,
  \ \
  \sigma_\psi^2 = \frac{1}{3}\int\frac{d^3k}{(2\pi)^3}\,k^{-2} P(k)\,,
    \label{sig0212psi2}
\end{eqnarray}
\end{subequations}
where we recall that $\sigma_0^2$ and $\sigma_1^2$ are the covariances
of the constraints $\Delta_0$ and $\vec\Delta_0$ defined in
Eqs.~(\ref{varcon}), and $W_R(k)$ is a smoothing window defined in
Eqs.~(\ref{smoothdel}) and (\ref{window}).  We use the Gaussian window
$W_R(k)=\exp(-k^2R^2/2)$.

To gain some insight into the significance of these results, let
us consider a proto-halo at high redshift.  If we suppose that the
gravity field is dominated by its small-scale components, we may
replace $\vec g_{\rm T}$ by $\vec g$. If we further ignore the
effects of initial constraints, in linear theory $D_r\sim D_t\sim
\frac{3}{2}H\sigma_v^2$ where $\sigma_v$ is the one-dimensional
velocity dispersion.  In one realization, the forces are
deterministic, but taken across the ensemble they are random and
fluctuating.  A diffusivity $D\sim H\sigma_v^2$ implies that these
fluctuations cause the typical particle to change its velocity by
order itself in about a Hubble time.  There is also a net drift
force $\vec A$ with a dissipative timescale of about $\delta^{-1}$
dynamical times, where $\delta$ is a typical magnitude of the
density fluctuations.  As a result, friction and diffusion will
significantly alter halos within a Hubble time of their collapse
at high redshift. We conclude that it is not valid to model halos
using the spherically symmetric Vlasov equation.  This conclusion
has also been reached previously by \cite{adc94}.

To summarize, Eqs.~(\ref{fplin})--(\ref{difflin}), and their more
detailed version, Eqs.~(\ref{adscaled})--(\ref{sumfun}), are the main
results of this subsection.  Eq.~(\ref{fplin}) has the standard form
of a Fokker-Planck equation.  The Fokker-Planck form arises naturally
and Eqs.~(\ref{driftlin})--(\ref{difflin}) are exact for Gaussian
fluctuations up to second order in perturbation theory. The diffusion
coefficients $\vec A$ and ${\bf D}$ are due to fluctuations in the
gravitational field over the ensemble of halos. The drift vector $\vec
A$ represents a source of gravitational acceleration in addition to
the acceleration $\vec g_{\rm T}$ produced by the mean matter
distribution.  The diffusion term $-{\bf D}\cdot\partial
f/\partial\vec v$ causes the temperature (i.e. velocity dispersion) to
increase (for a positive definite ${\bf D}$).

\subsection{Results for $\Lambda$CDM Model}

\begin{figure}
  \begin{center}
    \includegraphics[scale=0.8]{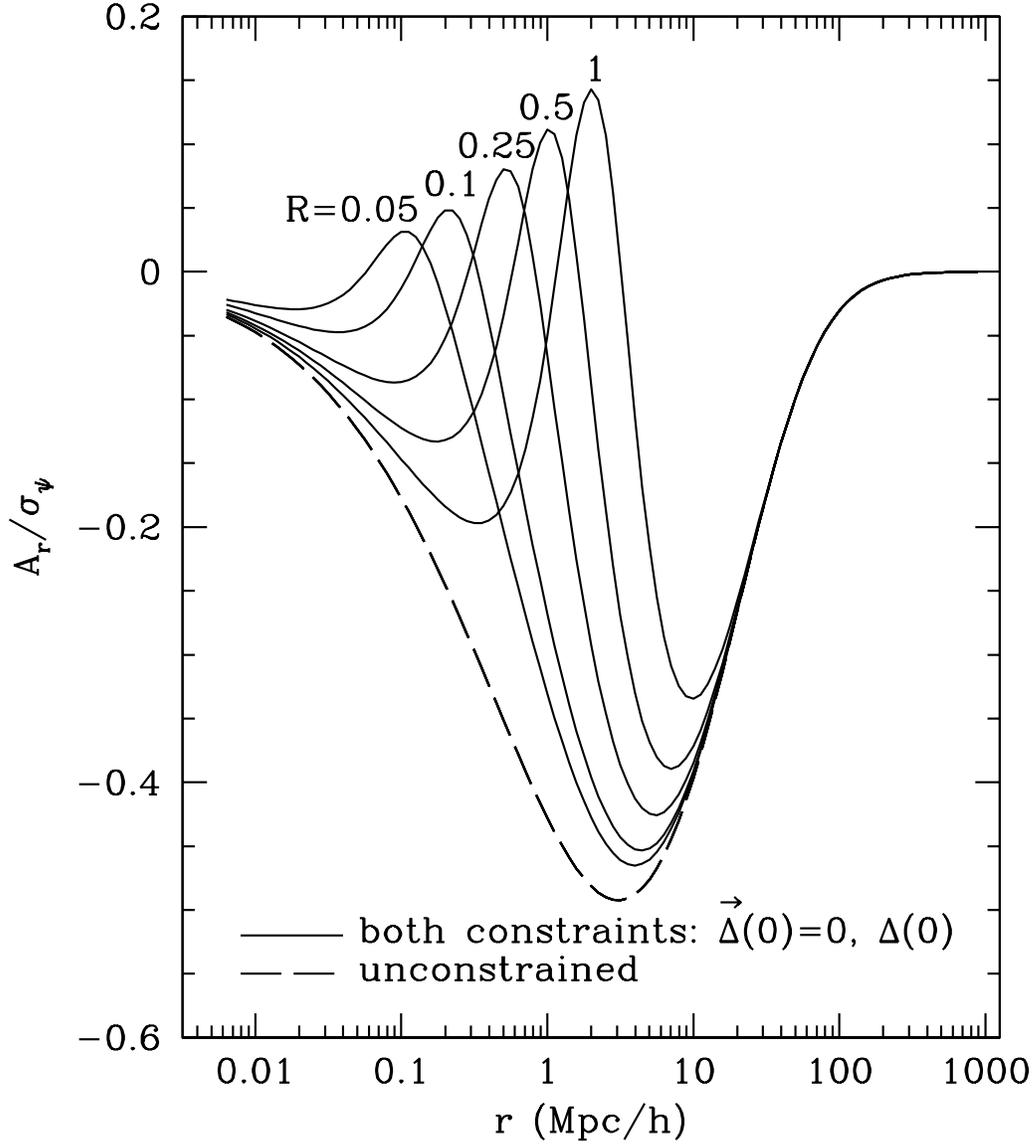}
  \end{center}
  \begin{quote}
    \caption{Radial drift coefficient $A_r$ plotted as a function of
    halo radius $r$ for Gaussian smoothing lengths $R=1,$ 0.5, 0.25, 0.1, and
    0.05 $h^{-1}$ Mpc.  It is computed from Eq.~(\ref{scala})
    for the $\Lambda$CDM model with $(\Omega_m,\Omega_\Lambda,
    \Omega_b,h)=(0.3,0.7,0.05,0.7)$.
    For comparison, the dashed curve shows
    $A_r$ for an unconstrained field.  The vertical axis shows the
    dimensionless $A_r/\sigma_\psi$, i.e. the drift
    normalized by the rms of the peculiar gravitational acceleration.}
    \label{fig:friction}
  \end{quote}
\end{figure}

\begin{figure}
  \begin{center}
    \includegraphics[scale=0.8]{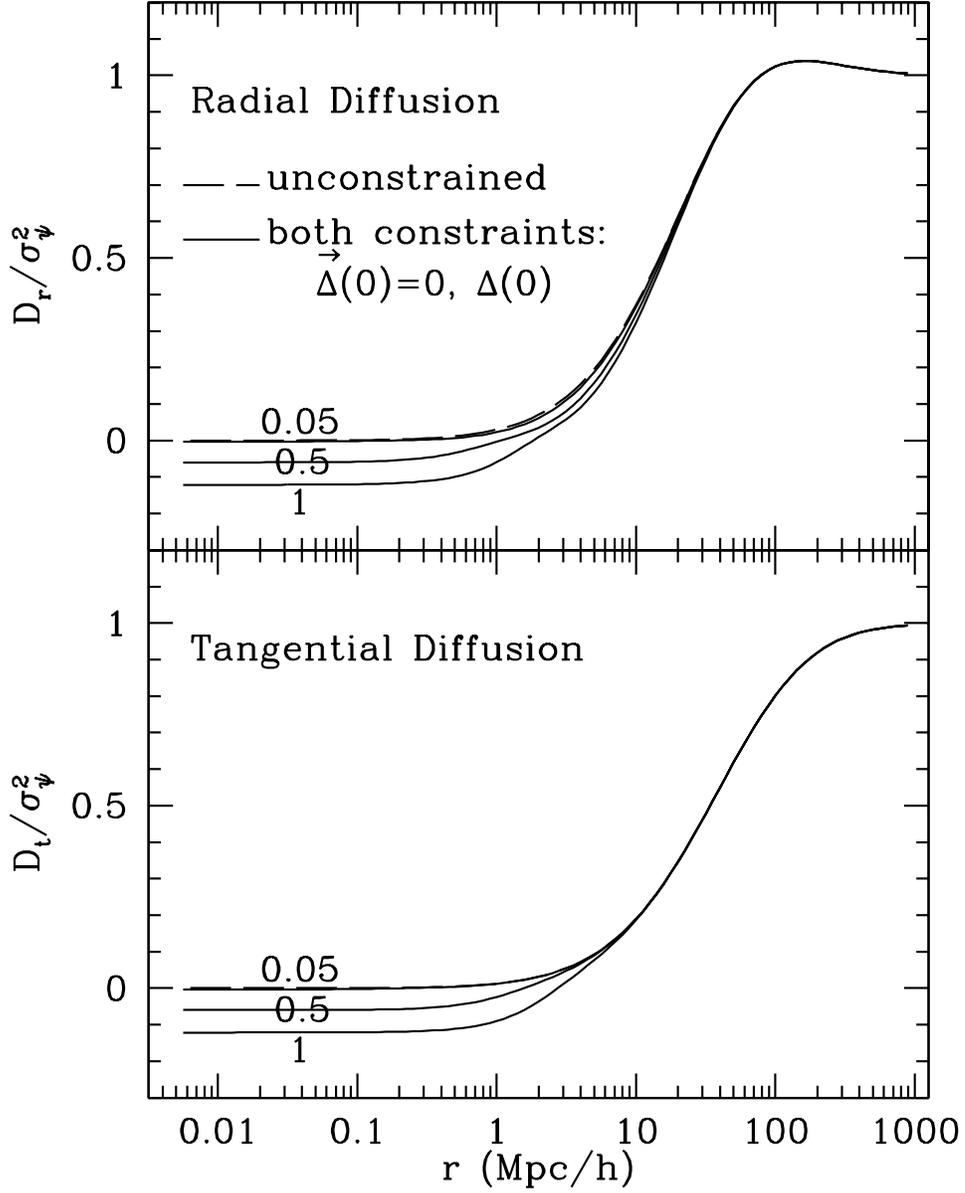}
  \end{center}
  \begin{quote}
    \caption{Diffusion coefficient $D$ in the radial ($r$) and
    tangential ($t$) directions plotted as a function of halo radius $r$ for
    Gaussian smoothing lengths $R=1$, 0.5, and 0.05 $h^{-1}$ Mpc.
    It is computed from Eqs.~(\ref{scaldr}) and (\ref{scaldt})
    for the same cosmological model as Fig.~1.
    The dashed curve in each panel is for an unconstrained field.
    The vertical axis shows the dimensionless $D_{ij}/\sigma^2_\psi$,
    i.e. the diffusivity normalized by the mean squared peculiar
    gravitational acceleration.}
    \label{fig:diffusion}
  \end{quote}
\end{figure}

Having derived analytic expressions for the diffusion coefficients in
the previous subsection, we now evaluate them for the currently
favored cosmological model at high redshift.
Figures~\ref{fig:friction} and \ref{fig:diffusion} show $A_r$, $D_r$,
and $D_t$ as a function of radius $r$ (measured from the halo center)
for a range of Gaussian smoothing lengths $R$ in $W_R(k)$.  We obtain
these results by numerically integrating Eqs.~(\ref{coeffs}) and
(\ref{sumfun}) with the linear power spectrum $P(k)$ for a flat
$\Lambda$CDM cosmological model with
$(\Omega_m,\Omega_\Lambda,\Omega_b,h)=(0.3,0.7,0.05,0.7)$.

The figures show a number of interesting features.  First, the
diffusion coefficients are negative over some regions of radius for
some parameters.  As we will show below, $\vec A$ is the gravitational
acceleration on particles in the halo caused by drift.  If the
friction were produced by the Chandrasekhar mechanism of tidal wakes,
one would expect $\vec A\parallel-\vec v$ (see \S~\ref{sec:globular}).
However, here $\vec A$ is independent of $\vec v$ and the particle
mass.  The source of friction is not the tidal wake of a flow past a
moving massive particle.  Instead it is the fluctuations of the
gravitational field of a Gaussian random process, and these
fluctuations can lead to a radially outward or inward force. We
discuss this further in \S~\ref{sec:discuss}.

More surprising at first is the fact that the diffusivities $D_r$
and $D_t$ can be negative.  Negative diffusivity causes the
velocity dispersion (or temperature) to decrease, indicating an
instability.  Such behavior is possible for gravitational systems.
In fact, this result is to be expected in the quasilinear regime
of gravitational instability because second-order perturbations
enhance gravitational collapse of dark matter halos
\citep{peeb80,bj94}. In the strongly nonlinear regime, after halo
virialization, we expect the diffusivities to become positive.

\begin{figure}
  \begin{center}
    \includegraphics[scale=0.8]{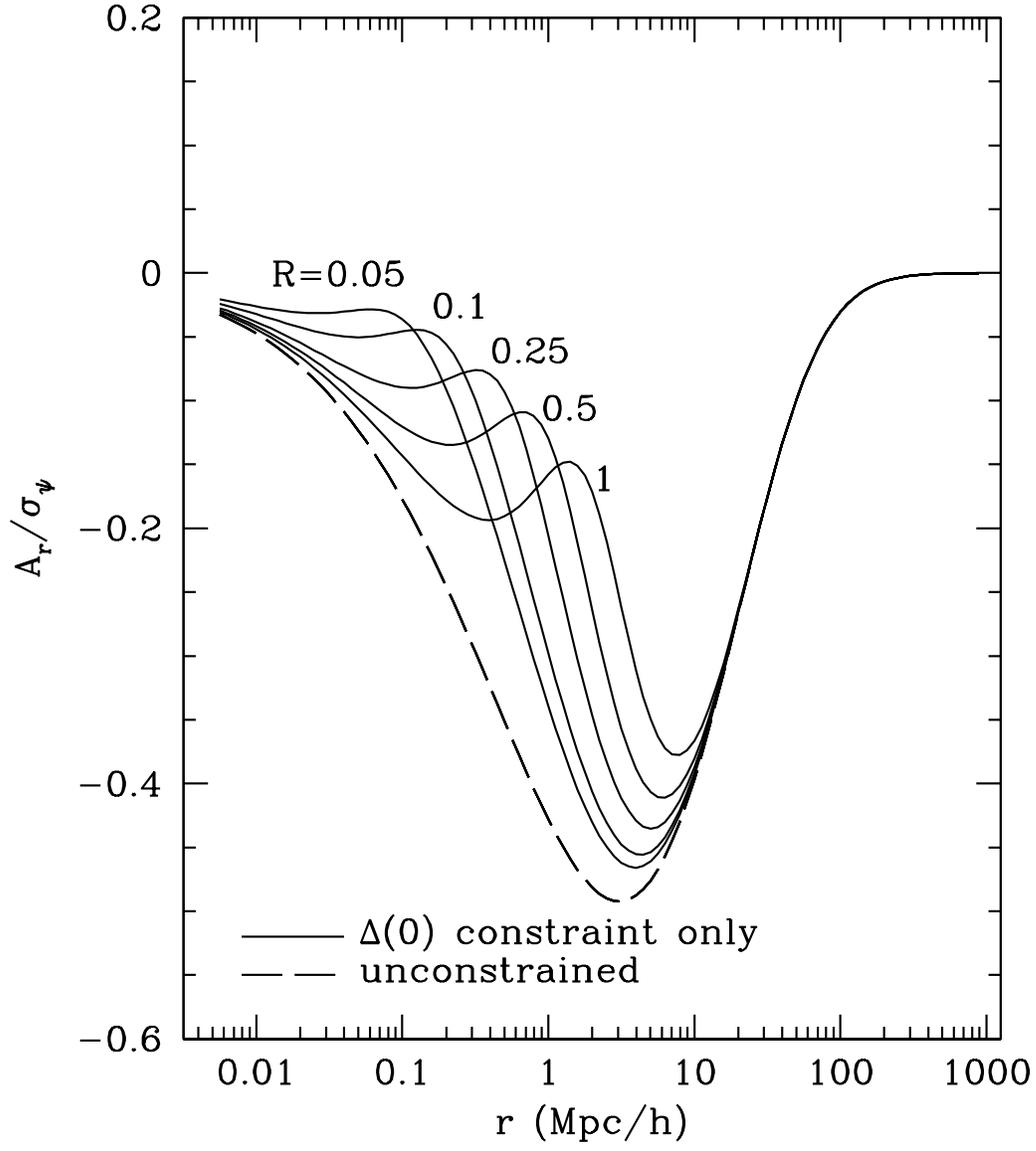}
  \end{center}
  \begin{quote}
    \caption{The same as Fig.~1 except that no extremum constraint
    (i.e., $\vec\Delta_0\equiv \vec\nabla\Delta(\vec x=0)=0$)
    is applied.}
    \label{fig:fig3}
  \end{quote}
\end{figure}

\begin{figure}
  \begin{center}
    \includegraphics[scale=0.8]{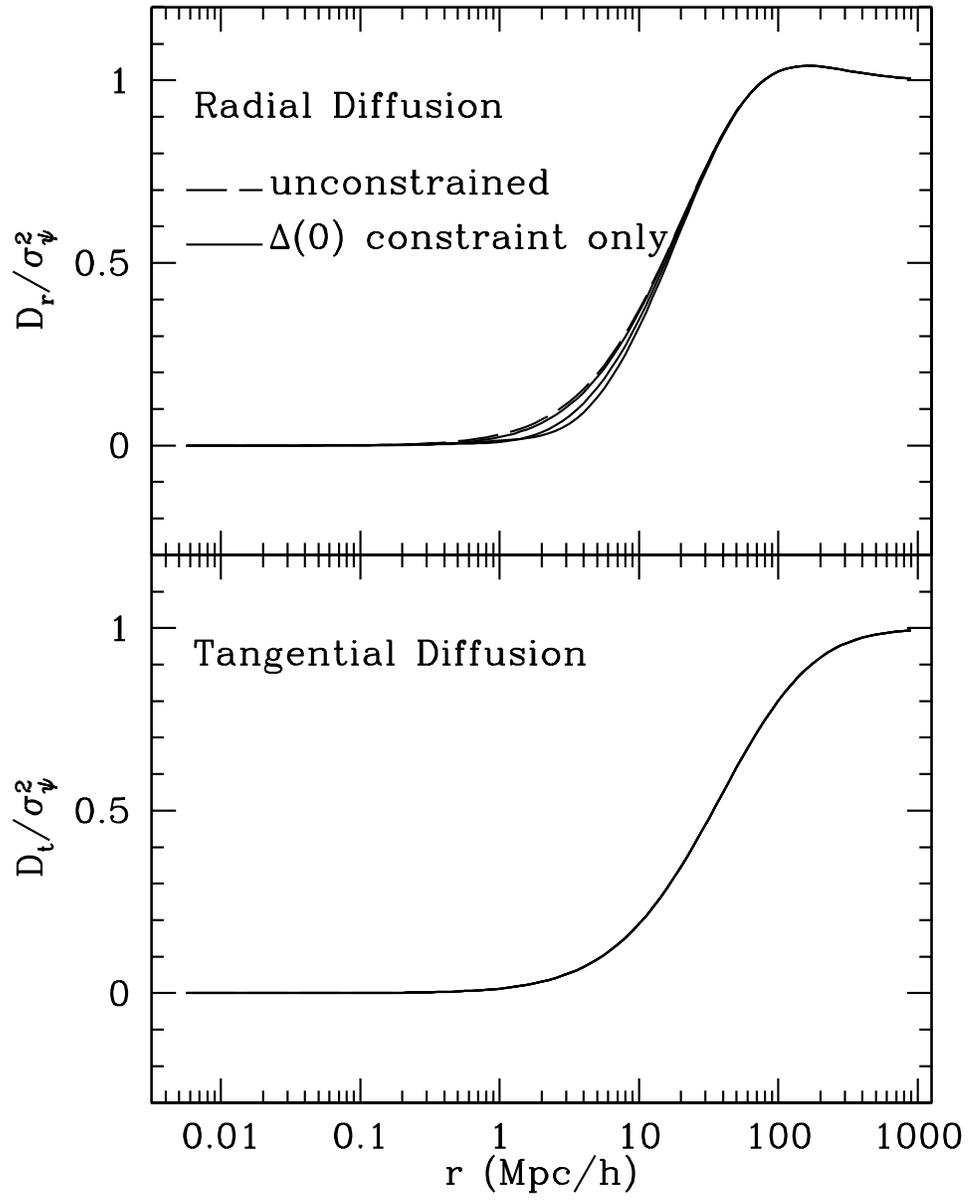}
  \end{center}
  \begin{quote}
    \caption{The same as Fig.\ 2 except that no extremum constraint
    is applied.}
    \label{fig:fig4}
  \end{quote}
\end{figure}

To elucidate the origin of the negative diffusivity, we show in
Figures~\ref{fig:fig3} and \ref{fig:fig4} that once the extremum
constraint $\vec\Delta_0\equiv \vec\nabla\Delta(0)=0$ is removed (by
setting $\sigma_1^{-2}=0$), we obtain $A_r\le 0$ and $D_r,D_t \ge 0$
for all radii.  The extremum constraint is therefore crucial to
generating positive $A_r$ and negative diffusivity seen in
Figures~\ref{fig:friction} and \ref{fig:diffusion}.  This effect is
easy to understand: if one stands at a random place in a Gaussian
field, nonlinear effects can speed up or slow down gravitational
collapse.  Gravitational instability, however, speeds up the evolution
in the vicinity of a density extremum.

Another obvious systematic effect with the diffusion coefficients in
Figures~\ref{fig:friction} and \ref{fig:diffusion} is their dependence
on the smoothing radius $R$ used to set the initial constraints.  For
the power spectrum assumed here, the maximum of $A_r$ occurs at about
$2R$.  In the limit $R\to0$, the variances $\sigma_0^2$ and
$\sigma_1^2$ diverge, causing the constraint terms in the diffusion
coefficients (i.e., those terms proportional to $\sigma_0^{-2}$ and
$\sigma_1^{-2}$) to vanish.

Figure~\ref{fig:diffusion} shows that the diffusivity is isotropic
at the halo center but shows radial anisotropy outside the center.
At small radii the tangential diffusivity is larger in absolute
value than the radial diffusivity.  Tangential diffusion gives the
infalling matter angular momentum.  Our treatment therefore
provides a new statistical description of the growth of angular
momentum by tidal torques \citep{w84} and may enable in principle
an explanation for the universal angular momentum profile found
recently by \cite{bull01}.

\subsection{Asymptotic Behavior}

It is instructive to work out analytically the asymptotic behavior
of the functions in Eqs.~(\ref{coeffs}) for small and large $r$.
For small $r$ we find,
\begin{subequations}\label{difcosmallr}
\begin{eqnarray}
  A_r(r,t) &=& r \left\{-\frac{1}{3}\sigma_\delta^2 + \alpha \bar\eta(0)
        + \frac{1}{1 - \bar\eta^2(0)/\sigma_1^2 \sigma_\psi^2}
        \left[\frac{3\bar\eta^2(0)}{\sigma_0^2} -
           \alpha \bar\eta(0) \right] \right\}
        + O(r^3) \\
  D_r(r,t) &=& -\left[\frac{\bar\eta(0)}{\sigma_1}\right]^2 + r^2 \left[
        \frac{1}{10}\sigma_\delta^2 + \frac{3}{5}\alpha \bar\eta(0)
         -\left(\frac{\bar\eta(0)}{\sigma_0} \right)^2 \right]
        + O(r^4) \ , \\
  D_t(r,t) &=& -\left[\frac{\bar\eta(0)}{\sigma_1}\right]^2 + r^2 \left[
        \frac{1}{30}\sigma_\delta^2
        + \frac{1}{5}\alpha \bar\eta(0) \right]
        + O(r^4) \ ,
\end{eqnarray}
\end{subequations}
where we have defined
\begin{subequations}
\begin{eqnarray}
   \sigma_\delta^2 &\equiv& \int\frac{d^3k}{(2\pi)^3}\,P(k) \ , \\
   \bar\xi(0) &=& 3\bar\eta(0) \equiv \int\frac{d^3k}{(2\pi)^3}\,P(k)W_R(k)\ ,\\
   \alpha & \equiv & \frac{ \int d^3k\, k^2 P(k)W_R(k)}
    {\int d^3k\, k^2 P(k)W_R^2(k)}
          = \frac{1}{3\sigma_1^2} \int\frac{d^3k}{(2\pi)^3}\, k^2 P(k)W_R(k) \ .
\end{eqnarray}
\end{subequations}

Eqs.~(\ref{difcosmallr}) show that if $\sigma_\delta^2$ is finite,
then as $r\rightarrow 0$, $A_r$ drops to zero while $D_r$ and
$D_t$ approach a negative constant as discussed above,
$-\bar\eta^2(0)/\sigma_1^2$.  This central value is decreased as
the smoothing scale is decreased (i.e. higher $\sigma_0$ and
$\sigma_1$) as shown in Figure \ref{fig:diffusion}. As $R\to0$, we
have $\sigma_0\to\sigma_\delta$ and $\sigma_1\to\infty$ as long as
$d\ln P/d\ln k>-5$ as $k\to\infty$.  In the limit
$\sigma_1\to\infty$ (no extremum constraint), $D_r$ and $D_t$ are
non-negative everywhere (cf.\ Figs.\ 2 and 4).  Thus, negative
diffusivity (instability) is produced by constraining the slope of
the density field. Constraints have the opposite effect on $A_r$:
as the comparison of Figures 1 and 3 confirms, the contributions
to $A_r$ from constraint terms are positive.

For large $r$, both $c_r(r)$ and $c_t(r)$ approach
$\sigma_\psi^2$, and $\gamma(r) \propto 1/r$ (assuming $P\propto
k$ as $k\to0$), so
\begin{equation}
        D_r(r,t) \rightarrow \sigma_\psi^2 \ ,\quad
        D_t(r,t) \rightarrow \sigma_\psi^2  \ .
\end{equation}
Figure \ref{fig:diffusion} shows that $D_r$ rises slightly above
$\sigma^2_\psi$ at a few hundred Mpc (for the standard model power
spectrum) before reaching $\sigma^2_\psi$.  This is because
$D_r=c_r - d\gamma/dr$ and $d\gamma/dr$ is negative for this range
of $r$. The drift $A_r(r,t)$ approaches zero at large $r$.  In
practice, the large-$r$ limit is unimportant because halo infall
occurs only from within about 10 Mpc.

\subsection{Power-law Models}

Another way to gain insight into the diffusion coefficients is to
examine their behavior for scale-free spectra $P(k)\propto k^n$.
Many of the integrals are analytic in this case.

We define a set of functions
\begin{equation}\label{Glab}
  G_l(a,b)\equiv\int_0^\infty x^{a-1}\,j_l(bx)\exp\left(-\frac{1}{2}
    x^2\right)\,dx\ .
\end{equation}
The integral converges for $a+l>0$.  Using properties of the
spherical Bessel functions $j_l(x)$, we obtain the following
useful relations:
\begin{subequations}\label{glrel}
\begin{eqnarray}
  bG_{l+1}(a,b)&=&(a+l-1)G_l(a-1,b)-G_l(a+1,b)\ ,
    \label{Glrecur1}\\
  bG_{l-1}(a,b)&=&(2+l-a)G_l(a-1,b)+G_l(a+1,b)\ ,
    \label{Glrecur2}\\
  G_0(a,0)&=&2^{(a-2)/2}\,\Gamma\left(\frac{a}{2}\right)\ .
    \label{Ggamma}
\end{eqnarray}
\end{subequations}
We also use the following relation, valid for $a+l>0$:
\begin{equation}\label{gla}
  g_l(a)\equiv\lim_{\sigma\to\infty}\int_0^\infty x^{a-1}
    \,j_l(x)\exp\left(-\frac{x^2}{2\sigma^2}\right)\,dx
    =2^{a-2}\sqrt{\pi}\,\Gamma\left(\frac{a+l}{2}\right)\,
    \left[\Gamma\left(\frac{3+l-a}{2}\right)\right]^{-1}\ .
\end{equation}

For Gaussian-tapered power-law spectrum
$P(k)=2\pi^2Bk^n\exp(-k^2R_0^2)$ and Gaussian smoothing
$W_R(k)=\exp(-k^2 R^2/2)$, Eqs.~(\ref{sumfun}) give, for $R_0\to0$
and $n>-3$,
\begin{subequations}\label{sumfunp}
\begin{eqnarray}
  r\eta(r) &=& Bg_1(n+2)\,r^{-(n+2)}\ ,\label{retap}\\
  \gamma(r) &=& Bg_1(n)\,r^{-n}\ \ \hbox{for}\ \ n > -1
    \ ,\label{gammap}\\
  \sigma_\psi^2-\frac{d\gamma}{dr}&=&
    B\begin{cases}
      ng_1(n)\,r^{-(n+1)}+\frac{1}{6}\,\Gamma\left(\frac{n+1}{2}
        \right)\,R_0^{-(n+1)}\ ,&n\ne-1\ ,\\
      -\frac{1}{9}+\frac{1}{6}{\cal C}+\frac{1}{3}\log(r/R_0)\ ,&n=-1\ ,
    \end{cases}
    \label{dr00}\\
  \sigma_\psi^2-\frac{\gamma}{r}&=&
    B\begin{cases}
      -g_1(n)\,r^{-(n+1)}+\frac{1}{6}\,\Gamma\left(\frac{n+1}{2}
        \right)\,R_0^{-(n+1)}\ ,&n\ne-1\ ,\\
      -\frac{4}{9}+\frac{1}{6}{\cal C}+\frac{1}{3}\log(r/R_0)\ ,&n=-1\ ,
    \end{cases}
    \label{dt00}\\
  \bar\xi(r) &=& BG_0(n+3,r/R)\,R^{-(n+3)}\ ,\ \
    r\bar\eta(r) = BG_1(n+2,r/R)\,R^{-(n+2)}\ ,\label{barxiretap}\\
  \sigma_0^2 &=& \frac{1}{2}B\,\Gamma\left(\frac{n+3}{2}
      \right)\,R^{-(n+3)}\ ,\ \
    \sigma_1^2 = \frac{1}{6}B\,\Gamma\left(\frac{n+5}{2}
      \right)\,R^{-(n+5)}\ ,\label{sig0212p}\\
  \sigma^2_\psi &=& \frac{1}{6}B\,\Gamma\left(\frac{n+1}{2}
    \right)\,R_0^{-(n+1)}\ \ \hbox{for}\ \ n > -1\ .
    \label{sigpsi2p}
\end{eqnarray}
\end{subequations}
Here, ${\cal C}=0.5772156649\cdots$ is Euler's constant.  For a
pure power-law spectrum with $n\le-1$, $\gamma(r)$ and
$\sigma_\psi^2$ diverge at long wavelength.  However,
Eqs.~(\ref{dr00}) and (\ref{dt00}) are valid for $n>-3$. For
$n\le-1$, $c_r^{-1}d\gamma/dr=1$ in Eq.~(\ref{scala}). For $n>-1$,
$\sigma_\psi^2$ converges at long wavelength but diverges at short
wavelength with $\sigma_\psi^2\propto R_0^{-(n+1)}$ as $R_0\to0$.

The limiting behavior of $\bar\xi(r)$ and $\bar\eta(r)$ are given
by
\begin{subequations}\label{limbarfun}
\begin{eqnarray}
  \bar\xi(r) &\approx&
    B\begin{cases}
      \frac{1}{2}\Gamma\left(\frac{n+3}{2}\right)\left(\frac{R}
      {\sqrt{2}}\right)^{-(n+3)}\left[1-\frac{(n+3)}{6}\frac{r^2}
      {R^2}+O(r^4)\right]\ ,&r\ll R\ ,\\
      -ng_1(n+2)\,r^{-(n+3)}\ ,&r\gg R\ ,
    \end{cases}
    \label{barxiap}\\
  \bar\eta(r) &\approx&
    B\begin{cases}
      \frac{1}{6}\Gamma\left(\frac{n+3}{2}\right)\left(\frac{R}
      {\sqrt{2}}\right)^{-(n+3)}\left[1-\frac{(n+3)}{10}\frac{r^2}
      {R^2}+O(r^4)\right]\ ,&r\ll R\ ,\\
      g_1(n+2)\,r^{-(n+3)}\ ,&r\gg R\ .
    \end{cases}
    \label{baretaap}
\end{eqnarray}
\end{subequations}
For all $r/R$, these functions obey the relation $\bar\xi+(R^2/r)
d\bar\xi/dr=-n\bar\eta(r)$.

\begin{figure}[t]
  \begin{center}
    \includegraphics[scale=1.0]{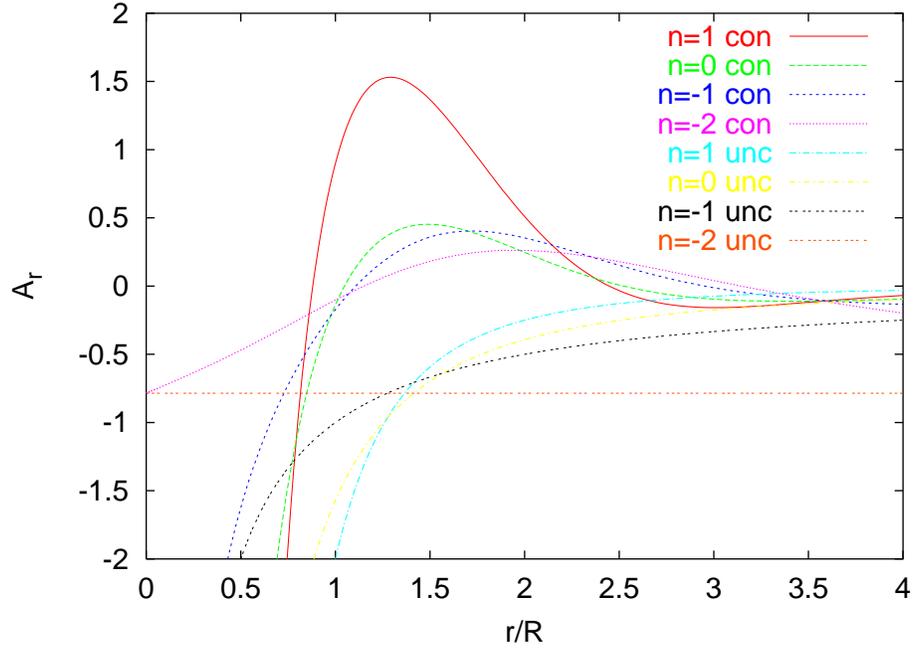}
  \end{center}
  \begin{quote}
    \caption{Radial drift coefficient $A_r(r)$ for power-law
    spectra, $P(k)\propto k^n$, plotted as a function of halo radius $r$
    scaled to the Gaussian smoothing radius $R$.
    Results are shown for four spectral indices, $n=-2,-1,0$, and 1,
    and for constrained and unconstrained fields.  The amplitude of
    $A_r$ corresponds to $P(k)=2\pi^2 B k^n$ normalized to $B=1$.}
    \label{fig:a-power}
  \end{quote}
\end{figure}

The drift coefficient is plotted in Figure \ref{fig:a-power} for
$n\in\{-2,-1,0,1\}$.  As expected from the previous section,
constraints make $A_r$ more positive.  For power-law spectra, however,
$\sigma_\delta^2$ diverges so one cannot use Eqs.~(\ref{difcosmallr})
to obtain the behavior as $r\to0$.  The exact results here show that
for $-2<n<4$, $A_r\to-\infty$ as $r\to0$. For $n<-2$, $A_r\to0$ as
$r\to\infty$.

These results show that models with lots of small-scale power and
substructure ($n>-2$ as $k\to\infty$) have a strong inward drift
force, while models that are smoother on small scales ($n<-2$)
have negligible drift force as $r\to0$.  This is easy to
understand qualitatively in the context of halo formation. Density
fields with lots of substructure have a nonspherical mass
distribution, with mass concentrated into subhalos.  For a given
halo, the mass between $r/2$ and $r$, for example, is concentrated
into overdense substructures whose gravity field on infalling
particles is stronger, on average, than if this mass were
uniformly distributed over a spherical shell as it is for the
average halo.  Given a density correlation function that rises
with decreasing $r$, there is a greater concentration of such
clumps interior to $r$ than outside of $r$, leading to a net
inward drift force.

\begin{figure}[t]
  \begin{center}
    \includegraphics[scale=1.0]{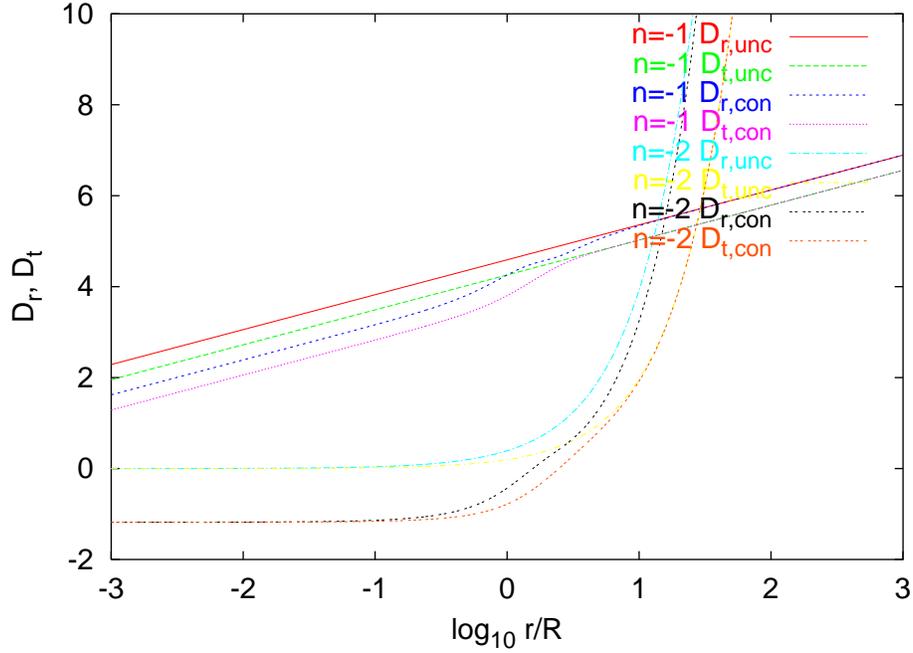}
  \end{center}
  \begin{quote}
    \caption{Diffusion coefficient $D$ in the radial ($r$) and
    tangential ($t$) directions for power-law spectra, $P(k)\propto k^n$,
    plotted as a function of halo radius $r$ scaled to the Gaussian smoothing
    radius $R$.  Results are shown for two spectral indices, $n=-1$ and $-2$,
    and for constrained and unconstrained fields.  For $n<-1$, constraints
    cause the diffusivity to approach a negative constant for
    $r\ll R$.  The amplitude of $D$ corresponds to
    $P(k)=2\pi^2 B k^n$ normalized to $B=1$.}
    \label{fig:d-power}
  \end{quote}
\end{figure}

Figure \ref{fig:d-power} shows the radial and tangential
diffusivity for power-law spectra with $n\le-1$.  For $n>-1$ the
diffusivity diverges because the velocity dispersion
$\sigma_\psi^2$ diverges as $k\to\infty$.  For $n=-1$ the
divergence is logarithmic resulting in a logarithmic dependence of
$D_r$ and $D_t$ on radius.  Figure \ref{fig:d-power} shows that
the effect of the constraints is to increase the diffusivity at
radii within a few smoothing lengths of the peak.  As we noted
above, constraints on the density gradient make the diffusivities
negative as $r\to0$.  For $n\le-1$ the diffusivities continue to
rise with increasing $r$ because $\sigma_\psi^2$ diverges as
$k\to0$.  For the physical power spectrum used in Figures
\ref{fig:friction} and \ref{fig:diffusion}, $\sigma_\psi^2$
converges at both small and large $k$, so the asymptotic behavior
is different.

\section{Discussion of the Fokker-Planck Equation}
\label{sec:discuss}

The use of the Fokker-Planck equation to describe dark matter
clustering is novel.  In order to gain insight into what this
approach may reveal, in this section we examine the Fokker-Planck
equation from several different perspectives.

\subsection{Velocity Moments}
\label{sec:velvom}

The Fokker-Planck equation describes relaxation processes in a
weakly collisional fluid.  In this regard it is similar to the
Boltzmann equation, in which the right-hand side of
Eq.~(\ref{boltzeq}) is replaced by a two-body collision integral
that is bilinear in the one-particle distribution function.  The
present situation is physically very different however ---
``collisions'' are not due to individual particles but rather to
the gravity field of density fluctuations in a Gaussian random
field.  Nevertheless, we can borrow one of the techniques used to
gain insight into the Boltzmann equation, viz. velocity moments,
to obtain a clear physical interpretation of the meaning of the
Fokker-Planck equation and the diffusion coefficients $\vec A$ and
${\bf D}$.

Moment equations are obtained by multiplying Eq.~(\ref{fplin}) by
powers of $v_i$ and integrating over velocity. The three lowest
moments give the mass density $\rho$, fluid velocity $\vec u$, and
velocity dispersion tensor $\theta_{ij}$:
\begin{subequations}\label{momdef}
\begin{eqnarray}
  \rho(\vec r,t) &=& \int d^3v\,f(\vec r,\vec v,t)
    \ ,\label{rhodef}\\
  u_i(\vec r,t) &=& \rho^{-1}\int d^3v\,f(\vec r,\vec v,t)\,v_i
    \ ,\label{udef}\\
  \theta_{ij}(\vec r,t) &=& \rho^{-1}\int d^3v\,f(\vec r,\vec v,t)
    (v_i-u_i)(v_j-u_j)\ .\label{thetadef}
\end{eqnarray}
\end{subequations}
Here we are using physical (proper) spatial coordinates rather
than comoving coordinates.  The density, velocity, and velocity
dispersion tensor obey the following equations:
\begin{subequations}\label{momeqs}
\begin{eqnarray}
  \frac{\partial\rho}{\partial t}+\frac{\partial}{\partial r_i}
    (\rho u_i)&=&0\ ,\label{continuity}\\
  \left(\frac{\partial}{\partial t}+u_k\frac{\partial}{\partial
    r_k}\right)u_i+\frac{1}{\rho}\frac{\partial}{\partial r_j}
    (\rho\theta_{ij})&=&g_{{\rm T}\,i}+A_i\ ,\label{euler}\\
    \left(\frac{\partial}{\partial t}+u_k\frac{\partial}{\partial
      r_k}\right)\theta_{ij}+\frac{\partial u_i}{\partial r_k}
      \theta_{jk}+\frac{\partial u_j}{\partial r_k}\theta_{ik}
      +\frac{1}{\rho}\frac{\partial}{\partial r_k}(\rho\theta_{ijk})
      &=&2D_{ij}\ .\label{heat}
\end{eqnarray}
\end{subequations}
The first equation is the usual fluid continuity equation.  The
second is the usual Euler equation, except that the pressure is
generalized to a matrix $\rho\theta_{ij}$ because the spatial
stress need not be diagonal for a collisionless fluid.  The
right-hand side has two force terms: the gravitational tidal field
$\vec g_{\rm T}$ produced by $\rho(\vec r,t)$, and the drift
acceleration $\vec A(\vec r,t)$ caused by the correlated
fluctuations of the mass density field.  The third equation
introduces a tensor $\theta_{ijk}$, which is defined exactly like
$\theta_{ij}$ except the integrand in Eq.~(\ref{thetadef})
acquires an additional factor $(v_k-u_k)$.  Eq.~(\ref{heat}) is a
generalized heat equation, which requires a little more
discussion.

The velocity dispersion tensor $\theta_{ij}$ generalizes the
temperature of an isotropic gas to the case of an anisotropic
velocity distribution. For an isotropic gas with
$\theta_{ij}=T\delta_{ij}$ (here $T$ has units of $v^2$) and
$D_{ij}=D\delta_{ij}$, this equation takes the more familiar form
\begin{equation}\label{isoheat}
  \frac{3}{2}\frac{dT}{dt}+(\vec\nabla\cdot\vec u\,)\,T
    +\frac{1}{\rho}\vec\nabla\cdot\vec q=3D\ ,
\end{equation}
where $q_i=\frac{1}{2}\rho\sum_{j=1}^3\theta_{ijj}$ is the heat
flux. The velocity gradient terms in Eq.~(\ref{heat}) generalize
the $pdV$ term (i.e., the $T\,\vec\nabla\cdot\vec u$ term) of
Eq.~(\ref{isoheat}) to an anisotropic velocity distribution.  For
a perfect gas, the heat flux term is absent because the Maxwellian
velocity distribution has vanishing skewness.  For an imperfect or
collisionless gas, however, the third moment of the velocity
distribution generates a generalized heat flux $\rho\theta_{ijk}$.
Notice that the heating rate is given not by the heat flux but
instead by its divergence.  For a weakly imperfect gas, the heat
flux can be approximated (e.g. in the Chapman-Enskog expansion
approach) by a conductive flux $\vec q=-\kappa\vec\nabla T$ where
$\kappa$ is the conductivity, but for a collisionless gas,
$\theta_{ijk}$ cannot be obtained simply from $\rho$, $\vec u$,
and $\theta_{ij}$.  Thus, Eqs.~(\ref{momeqs}) do not form a closed
system.  This is precisely why we must use a phase space
description for a collisionless gas instead of trying to modify
the fluid equations.  Eqs.~(\ref{momeqs}) and (\ref{isoheat}),
however, are pedagogically useful in showing us how to interpret
$\vec A$ and ${\bf D}$.

The velocity diffusivity tensor $D_{ij}$ appears as a new term in
the anisotropic heat equation, with no familiar counterpart in the
dynamics of a collisional gas.  Although $D_{ij}$ comes from a
{\it diffusion} (i.e., gradient) term in {\it phase space}, it
appears as a direct {\it local heating} term in {\it real space}
(i.e., coordinate space $\vec r\,$). Diffusion in velocity space
is equivalent to local heating: the temperature increases when
particles spread out in velocity at a given point in space.  The
diffusion occurs in velocity space not in real space.  The
presence of this term begs the question: where does the energy
come from?  From the exact kinetic equation (\ref{boltzeq}), it
follows that the heating term is given by the symmetrized first
velocity moment of the correlated force density $\vec F$ produced
by fluctuating gravitational fields.  From this we see that the
heating comes from gravitational energy. For this reason, it is
possible for $D$ to be positive or negative: gravitational
fluctuations can either locally heat or cool a gas.  One should
not try to draw conclusions about energy conservation from these
results because our calculations are performed in the accelerating
frame centered on a halo.  However, the equations of motion used
in deriving our kinetic theory imply local energy conservation.

\subsection{Langevin Equation}
\label{sec:langevin}

The Fokker-Planck equation describes the evolution of the probability
distribution for particles undergoing random walks.  Instead of
describing the system by a distribution function, we can specify
equations of motion for individual particle trajectories using the
Langevin equation \citep{langevin}.  Although this is reversing the
logic we used in deriving the Fokker-Planck equation, it is useful for
providing both an understanding and a Monte Carlo method for
numerically solving the Fokker-Planck equation.

The Langevin equation is a stochastic differential equation for
the individual particle trajectories. The Langevin equation
corresponding to Eq.~(\ref{fplin}) is given by the pair of
equations
\begin{equation}\label{langevin}
  \frac{d\vec r}{dt}=\vec v\ ,\ \
  \frac{d\vec v}{dt}=\vec g_{\rm T}(\vec r,t)+\vec A(\vec r,t)
  +\vec\Gamma(\vec r,t)\ ,
\end{equation}
where $\vec\Gamma$ is a force due to a zero-mean Gaussian random
process with covariance
\begin{equation}\label{stofor}
  \langle\Gamma_i(\vec r,t)\Gamma_j(\vec r,t')\rangle
  =2D_{ij}(\vec r,t)\,\delta_{\rm D}(t-t')\ .
\end{equation}
The velocity of a particle at time $t+dt$, $\vec{v}(t+dt)$, is
therefore given by a deterministic term $\vec{v}(t)+[\vec g_{\rm
T}(\vec r,t)+\vec A(\vec r,t)] dt$, plus a random term drawn from
a Gaussian distribution with variance $\propto D dt$. Subtle
differences exist between the cases where the variance $D$ is
evaluated at $(\vec{r}(t),t)$ (``Ito calculus'') vs. at
$(\vec{r}(t+dt/2),t+dt)$ (``Stratonovich calculus''). Details can
be found in \cite{risken}.  Note that the Langevin equation
requires $D>0$.

The interpretation of the Langevin equation is given by the
following result \citep{risken,gardiner}: If a sufficiently large
sample of particles is drawn at random at time $t_0$ from the
phase space distribution $f(\vec r,\vec v,t_0)$, and their orbits
are integrated using Eqs.~(\ref{langevin}), then at any later time
$t$ their positions and velocities are a sample from the phase
space density $f(\vec r,\vec v,t)$ obtained by integrating
Eq.~(\ref{fplin}).

Eqs.~(\ref{langevin}) are similar to the exact equations of motion
for individual particles given by Eqs.~(\ref{newton}), but there
are several important differences.  The gravity field $\vec g_{\rm
K}$ of one realization is replaced by the spherically symmetric
gravity field of the ensemble-average density profile. The effects
of spatially-varying forces in $\vec g_{\rm K}$ are represented by
the drift term $\vec A$ and the stochastic acceleration
$\vec\Gamma$. These additional terms have the effect that, at
least in the quasilinear regime (which we assumed in our
derivation of the Fokker-Planck equation), the phase space
evolution implied by the Langevin dynamics is equivalent to a
Monte Carlo sample from the phase space density.

This does not mean we have come full circle.  For individual
realizations of the ensemble, the density field is clustered and
not spherically symmetric and the phase space is six-dimensional.
The Langevin equations give orbits in a smooth, spherically
symmetric matter distribution. The latter is far easier to model
numerically, with much higher resolution possible, than the
original dynamics.  By taking advantage of the symmetries of the
average halo we have reduced the phase space from six dimensions
to three.

\subsection{Comparison with Classical Brownian Motion}
\label{sec:brownian}

The Fokker-Planck equation was first written almost a century ago
to describe Brownian motion \citep{fokker,planck}.  In this case,
a small macroscopic particle (a Brownian particle) undergoes a
random walk due to its collisions with individual molecules in a
liquid.  Instead of the phase space density of molecules, the
relevant quantity is the probability density $W(\vec v,t)$ for the
Brownian particle to have velocity $\vec v$ at time $t$. Assuming
a spatially homogeneous and isotropic medium, this probability
distribution obeys the Fokker-Planck equation
\begin{equation}\label{fpbrown}
  \frac{\partial W}{\partial t}=\frac{\partial}{\partial\vec v}
    \cdot\left(\gamma\vec v\,W+\gamma\frac{k_{\rm B}T}{m}
    \frac{\partial W}{\partial\vec v}\right)\ ,
\end{equation}
where $\gamma$ is a constant (given, for a small spherical body of
mass $m$ and radius $a$ immersed in a fluid of viscosity $\eta$,
by the Stokes formula $\gamma=6\pi\eta a/m$), $T$ is the
temperature, and $k_{\rm B}$ is the Boltzmann constant.
Eq.~(\ref{fpbrown}) has the same form as Eq.~(\ref{fplin}) if we
identify $\vec A=-\gamma\vec v$ for the drag term and
$D_{ij}=\gamma(k_{\rm B}T/m) \delta_{ij}$ for the diffusion term.

We note that the Brownian motion can be equally described by the
Langevin equations in Eqs.~(\ref{langevin}) and (\ref{stofor})
with $\vec{g}_T=0$, $\vec A=-\gamma\vec v$, and
$D_{ij}=\gamma(k_{\rm B}T/m)\delta_{ij}$.

Brownian motion describes a very different physical situation than
dark matter halo formation.  Why are both described by a
Fokker-Planck equation?  The reason is that both problems involve
random walks.  Let us examine this more closely.  In the case of
Brownian motion, the random walks arise from impulsive collisions
that cause the Brownian particle's velocity to change
significantly on a timescale $\gamma^{-1}$.  During this time
interval the particle moves a distance $\sim v/\gamma$.  After an
elapsed time $t$, the particle undergoes $\gamma t$ steps and
therefore moves a distance $\sim(v^2 t/\gamma)^{1/2}$.  In thermal
equilibrium, $v^2\sim k_{\rm B}T/m$ so the particle moves a
typical distance $\gamma^{-1}\sqrt{Dt}$ where $D=\gamma k_{\rm
B}T/m$ is the velocity diffusivity.  A random walk in space leads
naturally to a diffusion equation for the probability distribution
\citep{risken,gardiner}.

The dark matter case at first appears to be very different.
Instead of a large particle being buffeted by small ones, we are
studying small particles being buffeted by large mass fluctuations
(i.e. large in mass compared with the dark matter particles
themselves, whose mass never enters our discussion).  Moreover,
our derivation of the Fokker-Planck equation has assumed that the
motion is in the quasilinear regime described by
Eq.~(\ref{pertdv}) with a universal time-dependence for
$\vec\psi(\vec x,t)$. Therefore each particle has moved in a
straight line with (in appropriate units) a constant velocity,
which is not a random walk in space. However, {\it the velocity}
of a given particle is a random walk since $\vec\psi$ is obtained
by adding up the gravitational accelerations produced by all the
mass fluctuations in a Gaussian random field.  The net force on a
particle is a random walk at fixed time.

Although cosmic Gaussian random fields lead to a Fokker-Planck
equation, the diffusion coefficients are very different from
Brownian motion.  First, the drift acceleration $\vec A$ is
independent of velocity, in contrast with the viscous drag
acceleration $-\gamma\vec v$ for Brownian motion.  In the
quasilinear regime of gravitational clustering, the drift is not a
friction at all.  Instead it is radially directed (because of
spherical symmetry for the average halo), either inward or outward
depending on the density correlations of fluctuating substructure.
Second, the velocity diffusivity tensor $D_{ij}$ is not
proportional to the temperature; the dark matter is treated as
being completely cold with vanishing temperature.  In both cases
the diffusivity is proportional to the mean squared displacement.
In the cosmological case this depends simply on the power spectrum
of density fluctuations and not on the thermal velocity of the
particles.

In the Brownian case $\vec A$ and ${\bf D}$ are both proportional
to the Stokes drag coefficient $\gamma$.  The proportionality
between $\vec A$ and ${\bf D}$ is no accident but is a consequence
of the fluctuation-dissipation theorem, which states that these
coefficients have a proportionality given by the condition of
thermal equilibrium. It is instructive to see this in a
generalized case of Eq.~(\ref{fpbrown}).  Suppose that the
Brownian particle is moving in a time-independent potential
$\Phi(\vec r\,)$.  The Fokker-Planck equation then becomes
\begin{equation}\label{fpb}
  \frac{\partial W}{\partial t}+\vec v\cdot\frac{\partial W}{\partial
    \vec r}-\frac{\partial\Phi}{\partial\vec r}\cdot\frac{\partial W}
    {\partial\vec v}=\frac{\partial}{\partial\vec v}\cdot\left(\gamma
    \vec v\,W+\gamma\frac{k_{\rm B}T}{m}\frac{\partial W}{\partial
    \vec v}\right)\ .
\end{equation}
The stationary solution (for suitable boundary conditions) is
given by the Boltzmann distribution
\begin{equation}\label{equilsol}
  W=N\exp[-E/(k_{\rm B}T)]\ ,\ \
  E=\frac{1}{2}mv^2+m\Phi(\vec r\,)\ .
\end{equation}
where $N$ is a normalization constant.  With this solution not
only does the left-hand side of Eq.~(\ref{fpb}) vanish, but so
does the velocity flux on the right-hand side.  Drag and diffusion
balance each other in thermal equilibrium.  Note that the mean
squared velocity at fixed $\vec r$ equals $3k_{\rm B}T/m$. For a
distribution of masses of Brownian particles, thermal equilibrium
results in an equipartition of energy with the heaviest particles
moving most slowly.  This is a natural consequence of having
$(D/\gamma)^{1/2}$ equal the thermal speed.

In thermodynamic systems, the Fokker-Planck equation governs the
relaxation to thermal equilibrium.  In the cosmological case of
quasilinear clustering considered here, it is not thermal
equilibrium but rather the statistics of Gaussian random fields
that relate $\vec A$ and ${\bf D}$. In
Appendix~\ref{sec:stationary} we investigate our cosmological
Fokker-Planck equation for the relaxation to stationary solutions.

\subsection{Comparison with Globular Cluster Dynamics}\label{sec:globular}

Globular cluster evolution is the most studied application of the
Fokker-Planck equation in astrophysics.  Unlike the Brownian case,
gravity plays the dominant role here.  A star in a globular
cluster can be approximated as experiencing two types of
gravitational forces: a smoothly varying potential
$\Phi(\vec{r},t)$ due to the smoothed matter distribution in the
system, and a fluctuating force due to many two-body interactions
with other stars.  The phase space density of the stars obeys the
Fokker-Planck equation \citep{spitzer,bt88}
\begin{equation}\label{fp_globular}
  \frac{\partial f}{\partial t}+\vec v\cdot\frac{\partial f}{\partial
    \vec r} -\frac{\partial\Phi}{\partial\vec r}\cdot\frac{\partial f}
    {\partial\vec v}= -\frac{\partial}{\partial\vec v}\cdot
    \left[\vec A(\vec v\,)f \right]
    + \frac{1}{2} \frac{\partial^2}{\partial v_i \partial v_j}
    \left[ D_{ij}(\vec v\,)f \right] \,.
\end{equation}
This looks very similar to the cosmological kinetic equation
(\ref{fplin}) in the quasilinear regime, aside from a factor of
two in the diffusivity that is purely a matter of differing
conventions, and the velocity dependence of the diffusion
coefficients. However, the physics of $\vec A$ and $D_{ij}$ is
very different here.  The relaxation process in globular cluster
dynamics is {\it two-body relaxation} \citep{chandra}, i.e. the
dissipation arising from the fluctuating forces of discrete
Newtonian gravitating point masses.  Examining this case will shed
light on the relaxation processes that can affect galaxy halo
evolution.

The calculation of diffusion coefficients for two-body relaxation
is described in \S~8.3 of \cite{bt88}.  To understand the
results it is helpful to consider the process of Chandrasekhar
dynamical friction acting on a test particle with mass $m_t$ and
velocity $\vec v_t$ moving through a spatially homogeneous sea of
background particles of mass $m_b$ and velocity distribution
$f(\vec v_b)$ with mass density $\rho_b\equiv m_b\int f(\vec
v_b)\,d^3v_b$. For an isotropic Maxwellian distribution $f(\vec
v_b)\propto \exp(-v_b^2/2\sigma_b^2)$, the test body feels an
acceleration
\begin{equation}\label{dynfric}
  \frac{d\vec v_t}{dt}=-\frac{4\pi G^2\rho_b (m_t+m_b) \ln \Lambda}
    {v_t^3}\left[{\rm erf}(X) - \frac{2X}{\sqrt{\pi}} e^{-X^2} \right]
    \vec v_t\ ,
\end{equation}
where $X\equiv v_t/(\sqrt{2} \sigma_b)$.

Now treat each star within the globular cluster one at a time as a
test particle, with all the other stars serving as background
particles.  If the velocity distribution is Maxwellian and spatial
gradients are ignored, dynamical friction will contribute $\vec
A=-d\vec v/dt$ given by Eq.~(\ref{dynfric}) with $\vec v_t$
replaced by $\vec v$.  More generally, for a spatially homogeneous
distribution on scales larger than the interparticle spacing,
two-body relaxation gives \citep{bt88}
\begin{subequations}\label{ad}
\begin{eqnarray}
  \vec A &=& 4\pi G^2 m_b (m_t+m_b) \ln \Lambda \,
    \frac{\partial}{\partial \vec{v}} h(\vec{v}\,) \,,\label{a2b}\\
  D_{ij} &=& 4\pi G^2 m_b^2 \ln \Lambda \,
    \frac{\partial^2}{\partial v_i \partial v_j} g(\vec{v}\,) \,,
    \label{d2b}
\end{eqnarray}
\end{subequations}
where $\ln \Lambda$ is the Coulomb logarithm, and the Rosenbluth
potentials $h$ and $g$ are given by \citep{rosenbluth}
\begin{subequations}\label{rosenbluth}
\begin{eqnarray}
     h(\vec{v}\,) & \equiv & \int \frac{f(\vec{v}_b)\,d^3 v_b}
    {|\vec{v}-\vec{v}_b|}
    \,, \label{rosenh}\\
     g(\vec{v}\,) & \equiv & \int f(\vec{v}_b) |\vec{v}-\vec{v}_b|\,d^3 v_b \,.
     \label{roseng}
\end{eqnarray}
\end{subequations}
For the isotropic Maxwellian distribution $f(\vec v_b)\propto
\exp(-v_b^2/2\sigma_b^2)$, $\vec A=-\gamma(v)\vec v$ is a drag
force and the needed coefficients are
\begin{subequations}\label{ad2}
\begin{eqnarray}
  \gamma&=&\frac{4\pi G^2\rho_b (m_t+m_b) \ln \Lambda}{v^3}
    \left[{\rm erf}(X) - \frac{2X}{\sqrt{\pi}} e^{-X^2} \right]
    \ ,\\
  D_\parallel &=& 8\pi G^2 \rho_b m_b \frac{\sigma_b^2}{v^3} \ln \Lambda
    \left[ {\rm erf}(X) - \frac{2X}{\sqrt{\pi}} e^{-X^2} \right] \,,\\
  D_\perp &=& 8\pi G^2 \rho_b m_b \frac{1}{v} \ln \Lambda
    \left[ \left( 1- \frac{1}{2 X^2} \right) {\rm erf}(X)
    + \frac{1}{\sqrt{\pi}X} e^{-X^2} \right] \,,
\end{eqnarray}
\end{subequations}
where $X$ is defined as before (with $v_t\to v$), and $D_{ij}$ is
decomposed into two components:
\begin{equation}
    D_{ij} = \frac{v_i v_j}{v^2} D_\parallel +\frac{1}{2}
    \left( \delta_{ij} -\frac{v_i v_j}{v^2} \right) D_\perp \,.
\end{equation}
We note that the drift $\vec A$ contains two terms: the first term
is proportional to the test particle mass $m_t$ while the second
is independent of $m_t$.  The first term represents a
``polarization cloud'' effect, which arises from the fact that a
test particle moving relative to a background deflects more
upstream than downstream background particles
\citep{gilbert68,mulder83}.  This process results in a density
gradient and hence a drag force on the test particle, whose
amplitude increases linearly with $m_t$ since the gravitational
deflection is proportional to $m_t$.  In contrast, the second term
in $\vec A$ and the diffusivity tensor ${\bf D}$ in
Eq.~(\ref{ad2}) are both $\propto \rho_b m_b$ and independent of
$m_t$.  These two terms are due to fluctuations in the
gravitational potential caused by granularity in the background
particle distribution.

The derivation of the Fokker-Planck diffusion coefficients for
two-body relaxation is completely different from our derivation based
on quasilinear cosmological density fluctuations.  In both cases,
fluctuating forces lead to random walks in velocity space, hence to
the Fokker-Planck equation.  The nature of the fluctuating forces and
their consequences, however, are very different in the two cases.

Our cosmological Fokker-Planck derivation is exact (in
second-order cosmological perturbation theory); it yields no drag;
and the resulting drift and diffusion coefficients depend on
position but are independent of velocity, as shown in
Eqs.~(\ref{adscaled})--(\ref{sumfun}).  In contrast, the two-body
relaxation calculation is based a Taylor series expansion of the
phenomenological (not exact) master equation; it yields no radial
drift; and the drag and diffusion coefficients depend on velocity
but are (in the usual derivation) independent of position
\citep{rosenbluth}. Furthermore, the two-body relaxation rates are
proportional to $G^2$ while the rates we compute are proportional
to $G$ from Eq.~(\ref{adscaled}).  In the cosmology case, the
phase space correlations are built into the initial power spectrum
of density fluctuations, whose amplitude is given independently of
$G$. With two-body relaxation, however, correlations are
gravitationally induced by fluctuations instead of being present
ab initio, hence the rates pick up another factor of $G$.

The differences in the $\vec A$ term (radial drift versus drag)
have an important consequence for equipartition of energy. As
shown previously in the discussion of Brownian motion, drag and
diffusion work together to drive a system to thermal equilibrium
in which the mean squared velocity is proportional to $kT/m$.
Binney and Tremaine show (p.\ 513 of Binney \& Tremaine 1988) show
that this also happens with two-body relaxation. Because
equilibrium systems with only gravitational forces generally have
negative specific heat, there is no stable thermal equilibrium.
Nonetheless, two-body relaxation tends to drive the velocity
distribution toward the Maxwellian form, with equipartition of
different mass species.

In the quasilinear cosmological case, the drift and diffusion
coefficients are independent of both velocity and particle mass.
Therefore the equilibrium velocity distribution cannot depend on
particle mass --- there is no equipartition of energy.  This
result is similar to violent relaxation, a process arising during
the initial collapse and virialization of a dark matter halo when
the gravitational potential is rapidly varying in time
\citep{dlb67}.  Our relaxation process, however, arises in the
quasilinear regime and is present even if $\partial\phi/\partial
t=0$ as it is (at fixed comoving position) if $\Omega_{\rm m}=1$.
Moreover, violent relaxation (and its partner, phase mixing) are
generally understood as arising from the collisionless dynamics
represented by the left-hand side of Eq.~(\ref{fp_globular}),
while both two-body relaxation and our cosmological relaxation
process are represented by the ``collision'' terms on the
right-hand side.

Why did our cosmological calculation not yield a drag term (and a
corresponding diffusion term linked to it by the
fluctuation-dissipation theorem)?  We suspect that it is because
our calculation was limited to small-amplitude perturbations about
a homogeneous and isotropic expanding cosmological model.  The
induced effects of substructure would only come in at higher order
in perturbation theory.  In the nonlinear regime we expect two
types of drift terms to be present: $\vec A=\alpha\hat
r-\gamma\vec v$, where $\alpha$ is the radial drift and $\gamma$
is the drag coefficient.  The Chandrasekhar calculation suggests
that the drag and its accompanying diffusivity will depend on both
position (through $\rho_b$) and velocity.  The radial drift is
absent in the Chandrasekhar calculation because the background
there is assumed to be spatially homogeneous; with an
inhomogeneous distribution in a virialized halo there should be a
radial drift (and corresponding diffusivity) that could depend on
both position and velocity.  In the nonlinear regime drag should
arise from dynamical friction just as with the globular clusters,
except that the ``background'' particles whose discreteness causes
the $m_b$ terms in Eqs.~(\ref{ad2}) are now the subhalos and
substructures that rain upon a halo.   Because we have not yet
extended our derivation to the fully nonlinear regime, these
expectations, while plausible, have yet to be demonstrated.

\section{Summary and Conclusions}
\label{sec:conclu}

In this paper we have developed a cosmological kinetic theory, valid
to second-order in perturbation theory, to describe the evolution of
the phase-space distribution of dark matter particles in galaxy halos.
This theory introduces a new way to model the early phases of galaxy
halo formation, which has traditionally been studied by analytic
infall models or numerical N-body methods.

The key physical ingredients behind our kinetic description are
stochastic fluctuations and dissipation caused by substructures
arising from a spectrum of cosmological density perturbations. The
kinetic equation that we have obtained at the end of our
derivation, Eqs.~(\ref{fplin})--(\ref{difflin}), has the standard
form of a Fokker-Planck equation.  The diffusion coefficients
$\vec{A}$ and {\bf D} represent acceleration due to a drift force
and velocity-space diffusion, respectively.  To second order in
perturbation theory, these coefficients are related to various
covariance matrices of the cosmological density and velocity
fields given by Eqs.~(\ref{driftlin})--(\ref{difflin}).  These
expressions can in turn be written in terms of the familiar linear
power spectrum $P(k)$ of matter fluctuations, as shown in
Eqs.~(\ref{coeffs}) and (\ref{sumfun}).

The results for $\vec{A}$ and {\bf D} from second-order
perturbation theory are shown in Figures~1--4 for the currently
favored $\Lambda$CDM model, and in Figures~5 and 6 for various
scale-free models with power-law $P(k)\propto k^n$. Our results
indicate that dissipative processes are important during the
initial collapse of a dark matter halo, and it is not valid to
model halos using the spherically symmetric Vlasov (collisionless
Boltzmann) equation. Furthermore, we find that the diffusivity is
initially negative close to the center of a halo, indicating a
thermodynamic instability.  The source of this instability is
gravity: perturbations enhance gravitational collapse of dark
matter halos.

We emphasize that our derivation leading to the Fokker-Planck
equation given by Eqs.~(\ref{fplin})--(\ref{difflin}) is exact to
second order in cosmological perturbation theory.  We do not
follow the frequently used approach (e.g. Binney \& Tremaine 1988)
that begins by expanding the collisional terms of the non-exact
master equation in an infinite series of changes in phase-space
coordinates (i.e. $\Delta \vec{w}$), and arrives at the
Fokker-Planck equation by assuming weak encounters (i.e. small
$|\Delta \vec{w}|$) and truncating the series after the
second-order terms.  Instead, our starting point is
Eq.~(\ref{boltzeq}), the first BBGKY hierarchy equation for a
smooth one-particle distribution function.  This is an exact
kinetic equation; solving it requires specifying the one-particle
distribution at an initial time and the two-particle correlation
function at all times.  We have shown in \S~\ref{sec:dfgauss} and
Appendix~\ref{sec:pgauss} that if the cosmological density field
$\delta$ and displacement (or velocity) field $\psi$ are Gaussian
random fields, the resulting two-particle correlation function
depends on the one-particle distribution in a simple way given by
Eq.~(\ref{f2cexact}). Eq.~(\ref{f2cexact}) is exact to second
order in perturbation theory and provides the closure relation for
the BBGKY hierarchy in Eq.~(\ref{boltzeq}).  Combining
Eq.~(\ref{f2cexact}) with Eq.~(\ref{boltzeq}) results in
Eq.~(\ref{fplin}), which is a Fokker-Planck equation.

The cosmological dissipative processes identified in this paper
are quite different from the standard two-body relaxation, which
also leads to a Fokker-Planck equation.  We find that there is no
dynamical friction (i.e. terms proportional to $-\vec{v}$ in the
diffusion coefficients) in second-order cosmological perturbation
theory. Instead, there is a radial drift toward (or away from) the
center of a halo due to the clustering of matter within the halo.
The usual treatment of two-body relaxation has no such drift term
because it assumes the medium to be homogeneous.  Thus we have
identified a new relaxation process that must affect the
phase-space structure of dark matter halos.

Although our derivation is exact to second order in the density
fluctuations, it is valid only when the fluctuations are small. We
are therefore describing only the early stages of dark matter halo
formation.  The expressions we obtain for the diffusion
coefficients are plainly invalid in the nonlinear regime, where we
expect dynamical friction to be present as well as radial drift.
In a later paper we will investigate the nonlinear generalizations
of the exact second-order results derived in this paper.

The Fokker-Planck description should still apply in the nonlinear
regime provided that the fluctuating force on a particle can be
modelled as a Markov process consisting of a random walk of many
steps.  We conjecture that this description will be approximately
valid when the matter distribution is modelled as a set of clumps
(i.e., the halo model) which scatter individual dark matter
particles away from the orbits they would have in a smooth,
spherical potential. The Fokker-Planck description should apply
not only to the initial collapse and virialization of a dark
matter halo containing substructure, but also to the subsequent
evolution under minor mergers.  Our inability to solve exactly the
nonlinear dynamics, however, means that the diffusion coefficients
will have to be calibrated using the nonlinear halo model or
numerical N-body simulations.

Calibrating the diffusion coefficients is not equivalent to
reproducing the N-body simulations.  N-body simulations provide a
Monte Carlo solution of the BBGKY hierarchy, which is much more
general than the Fokker-Planck equation.  If the simulation
results are describable by Fokker-Planck evolution, this already
represents a significant new result.  Moreover, even if functional
forms for the radial dependence of the diffusion coefficients have
to be calibrated with simulations, the solution of the
Fokker-Planck equation contains far more information because it
gives the complete phase-space density distribution, not simply
radial profiles.

Another important step will therefore be to actually solve the
Fokker-Planck equation.  In deriving it we have identified and
quantified the relaxation processes affecting halo formation and
evolution, but this is only the first step.  In
Appendix~\ref{sec:stationary} we presented solutions in
unrealistically simple cases just to highlight the roles played by
radial drift, drag, and diffusion. An outstanding question is
whether these processes actually are rapid enough to erase the
memory of cosmological initial conditions sufficiently so that
dark matter halos relax to a universal profile.  Once we have a
nonlinear model for the diffusion coefficients, we will address
this important question.

The work presented in this paper has a number of implications.
First, we have highlighted the importance of the phase space
density for understanding dark matter dynamics, and have developed
a method for calculating its evolution during the early phases of
structure formation.  Second, we have identified a dissipative
process that may erase memories of initial conditions during the
early phases of galaxy halo evolution.  This erasure is necessary
for universal halo density profiles, although further
investigation is needed to quantify how this dissipative process
affects the density profiles.  Third, the statistical description
of the potential fluctuations caused by substructure that we have
developed can also be added to investigate related problems such
as the heating of galactic disks \citep{benson03} and the merging
of central black holes when galaxy halos merge
\citep{kh00,hk02,hb03}.

\acknowledgments

We thank Jon Arons, Avishai Dekel, and Martin Weinberg for useful
conversations.  The Aspen Center for Physics provided a stimulating
environment where a portion of the research was carried out.  C.-P. M
is partially supported by an Alfred P. Sloan Fellowship, a Cottrell
Scholars Award from the Research Corporation, and NASA grant
NAG5-12173.

\appendix
\section{Statistics of Constrained Gaussian Random Fields}
\label{sec:pgauss}

In this Appendix we derive expressions for the probability
distributions of displacement and density perturbation,
$p(\vec\psi_1)$ and $p(\delta_2,\vec\psi_1)$, that are needed in
Eqs.~(\ref{f1p1}) and (\ref{f2cp1}).  We then use them to evaluate
Eqs.~(\ref{f1p1}) and (\ref{f2cp1}) at high redshift when the
matter distribution was described by a Gaussian random field of
small-amplitude density perturbations.  We use comoving spatial
coordinates $\vec x=\vec r/a(t)$.

In accordance with the standard cosmological model, we assume that
$\delta(\vec x,t)$ is a Gaussian random field with power spectrum
$P(k,t)$, and that the displacement and density are related by
Eq.~(\ref{linpsi}).  These are the appropriate assumptions at high
redshift in standard models of structure formation. Strictly speaking,
at high redshift, when linear theory applies, the evolution of the
distribution function is trivial because the density and velocity
fields evolve by spatially-homogeneous amplification. Here we use the
statistics of Gaussian random fields to compute quantities driving the
lowest-order corrections to linear evolution.  We will find that the
fluctuating force is formally second order in perturbation theory.

A Gaussian random field is described most simply by its Fourier
transform, which we define by
\begin{equation}
  \label{fourier}
  \delta(\vec x\,)=\int \frac{d^3k}{(2\pi)^3}\,\delta(\vec k\,)
  e^{i\vec k\cdot\vec x}\ .
\end{equation}
We suppress the time dependence for convenience.  Before we apply a
constraint, both $\delta(\vec x\,)$ and $\delta(\vec k\,)$ have zero
mean.  The covariance of $\delta(k)$ gives the power spectrum,
\begin{equation}
  \label{pspect}
  \left\langle\delta(\vec k_1)\delta(\vec k_2)\right\rangle=
    (2\pi)^3 P(k_1) \delta_{\rm D}(\vec k_1+\vec k_2)\,,
\end{equation}
where $\delta_{\rm D}$ is the Dirac delta function.  In the linear
regime, $\vec\psi(\vec x\,)$ is a linear functional of
$\delta(\vec x\,)$.  The multivariate distribution of any linear
functional of a Gaussian random field is itself Gaussian (i.e.
multivariate normal) and is therefore described completely by its
mean and covariance matrix.  This remains true even for a
constrained Gaussian random field \citep{bbks}.

In our case, we wish to constrain the initial density field to
represent a proto-halo. The constraints are applied to the
smoothed density field
\begin{equation}
  \label{smoothdel}
  \Delta(\vec x\,)=\int d^3x'\,W_R(\vec x-\vec x^{\,\prime}\,)
    \delta(\vec x^{\,\prime}\,)=\int\frac{d^3k}{(2\pi)^3}\,
    W_R(\vec k\,)\delta(\vec k\,)e^{i\vec k\cdot\vec x}\ ,
\end{equation}
where
\begin{equation}
  \label{window}
  W_R(\vec x\,)=\int\frac{d^3k}{(2\pi)^3}\,W_R(\vec k\,)
    e^{i\vec k\cdot\vec x}
\end{equation}
is a smoothing window with unit volume integral, $\int
d^3x\,W_R(\vec x\,)=1$.  The subscript $R$ indicates the smoothing
length.  We assume that the smoothing window is spherically
symmetric so that $W_R(k)$ is real and depends only on the
magnitude of the wavevector.  Throughout the paper we will use the
Gaussian window $W_R(x)=(2\pi R^2)^{-3/2}\exp(-x^2/2 R^2)$ and
$W_R(k)=\exp(-k^2 R^2/2)$.

Unfortunately there is no analytic theory to tell us where halos will
form.  However, there are a variety of plausible approaches.  For
example, following \cite{bbks} we might constrain $\Delta$ to have a
maximum of specified height, orientation, and shape at $\vec
x=0$. This peak constraint involves ten variables ($\Delta$ and the
components of its first and second derivatives with respect to each of
the coordinates), presenting a formidable challenge to evaluating the
covariance matrix of the four variables $(\delta_2,\vec\psi_1)$.  The
peak constraint is appealing but it is difficult to calculate and in
practice it has not been found to predict well the actual formation
sites of dark matter halos \citep{kqg}.

Instead of trying to impose complicated constraints for halo
formation (e.g. Monaco et al.\ 2002), we err on the side of
simplicity by choosing to use only the zeroth and first
derivatives of the smoothed initial density field $\Delta(\vec
x\,)$.  We require that $\vec x=0$ be an {\it extremum} of the
smoothed density field by requiring $\vec\nabla\Delta(0)=0$.  Of
course, an extremum may be a maximum, a minimum, or a saddle
point.  We therefore keep track also of the smoothed density
$\Delta_0\equiv\Delta(0)$. Provided that $\Delta_0$ is greater
than 2 standard deviations, the extrema of $\Delta(\vec x)$ are
close to the peaks \citep{bbks}.

How will we know if our results depend strongly on the choice of
constraint? We will do this in two ways.  First, we will vary the
height of the extremum, $\Delta_0$.  Second, we will drop the
gradient constraint and use the single constraint $\Delta_0$.
While these tests will not prove that we have a good model for the
sites of halo formation, they do provide control cases for us to
assess the sensitivity of our results to the details of the
initial constraints.

Now we must calculate the mean and covariances of
$(\delta_2,\vec\psi_1)$ subject to constraints $(\Delta_0,\vec\Delta_0)$
where
\begin{equation}
   \label{delta0}
   \Delta_0\equiv \Delta(\vec x=0)\,,\qquad
    \vec\Delta_0\equiv \vec\nabla\Delta(\vec x=0) \,.
\end{equation}
The method for this calculation is based on the theorem presented in
Appendix D of \cite{bbks}, which states that if $Y_A$ and $Y_B$ are
zero-mean Gaussian variables (more generally, vectors of any length),
then the conditional distribution for $Y_B$ given $Y_A$, $p(Y_B\vert
Y_A)=p(Y_A,Y_B)/p(Y_A)$, is Gaussian with mean
\begin{equation}
  \label{meany}
 \left\langle Y_B\vert Y_A\right\rangle\equiv\int dY_B\,p(Y_B\vert
   Y_A)\,Y_B=\left\langle Y_B\otimes Y_A
    \right\rangle\left\langle Y_A\otimes Y_A\right\rangle^{-1}Y_A
\end{equation}
and covariance matrix
\begin{equation}
  \label{covy}
  C(Y_B,Y_B)\equiv\int dY_B\,p(Y_B\vert Y_A)\,\Delta Y_B\otimes
    \Delta Y_B=
    \left\langle Y_B\otimes Y_B\right\rangle-\left\langle Y_B\otimes Y_A
    \right\rangle\left\langle Y_A\otimes Y_A\right\rangle^{-1}
    \left\langle Y_A\otimes Y_B\right\rangle\ ,
\end{equation}
where $\Delta Y_B=Y_B-\left\langle Y_B\vert Y_A\right\rangle$. The
symbol $\otimes$ denotes a tensor product and angle brackets
denote mean values.  Thus, $\left\langle Y_A\otimes
Y_A\right\rangle$ is the covariance matrix of the constraints
while $C(Y_B,Y_B)=\langle\Delta Y_B\otimes\Delta Y_B\rangle_C=
\langle\Delta Y_B\otimes\Delta Y_B\vert Y_A\rangle$ is the
covariance matrix of $Y_B$ subject to the constraints.  (The
symbol $C$, used either as a function or as a subscript, implies
that a constraint is applied.)  The juxtaposition of two matrices
implies matrix multiplication.

\subsection{Covariance of Constraints and Variables}

To apply these results, first we compute the covariance matrix of
constraints $Y_A=(\Delta_0,\vec\Delta_0)$.  Using
Eqs.~(\ref{fourier})--(\ref{window}), we get the covariances
\begin{subequations}\label{varcon}
\begin{eqnarray}
    \l \Delta_0^2 \r  &=& \sigma_0^2 \equiv\int\frac{d^3k}{(2\pi)^3}\,
    P(k)W_R^2(k)\ ,  \label{var00}\\
    \l \Delta_0 \vec\Delta_0 \r &=& 0 \ ,\label{var0i}\\
        \l \vec\Delta_0 \otimes \vec\Delta_0 \r &=&  \sigma_1^2 {\bf I}\ ,
    \quad \sigma_1^2 \equiv \frac{1}{3}
    \int \frac{d^3k}{(2\pi)^3}\,k^2P(k)W_R^2(k)\ .\label{varij}
\end{eqnarray}
\end{subequations}
where ${\bf I}$ is the unit tensor.

Next we consider the covariance matrix of the variables
$Y_B=(\delta_1, \delta_2,\vec\psi_1)$.  Since
$\delta=-(\partial/\partial\vec{x})\cdot \vec\psi$, the
displacement field $\vec\psi$ has an additional factor $i\vec
k/k^2$ in the integrand of Eq.~(\ref{fourier}).  For the
off-diagonal elements of the covariance matrix of $Y_B$ we obtain
\begin{subequations}\label{covyb}
\begin{eqnarray}
  \left\langle\delta_1\delta_2\right\rangle &=& \xi(r)\equiv\int
    \frac{d^3k}{(2\pi)^3}\, P(k)j_0(kr)\ ,\quad\quad
    \vec r\equiv\vec x_1-\vec x_2\ ,\label{cov12}\\
  \left\langle\delta_1\vec\psi_1\right\rangle &=& 0\ , \label{cov1i}\\
  \left\langle\delta_2\vec\psi_1\right\rangle &=& -\eta(r)\vec r\ ,\quad
    \eta(r)\equiv\int\frac{d^3k}{(2\pi)^3}\,P(k)\frac{j_1(kr)}{kr}\ ,
    \label{cov2i}
\end{eqnarray}
\end{subequations}
where the spherical Bessel functions are $j_0(x)=x^{-1}\sin x$ and
$j_1(x)=x^{-2}(\sin x-x\cos x)= -dj_0/dx$.  The diagonal variances
are
\begin{subequations}\label{covybb}
\begin{eqnarray}
  && \left\langle\delta_1^2\right\rangle =
  \left \langle\delta_2^2\right\rangle = \xi(0)\ , \label{covybbs}\\
  && \left\langle \vec\psi_1 \otimes\vec\psi_1\right\rangle =
    \sigma_\psi^2{\bf I}\ , \quad
    \sigma_\psi^2=\frac{1}{3}\int\frac{d^3k}{(2\pi)^3}\, k^{-2}P(k)\ .
    \label{covybbt}
\end{eqnarray}
\end{subequations}
Note that in Eqs.~(\ref{covyb}) (and in the following) we are
using $\vec r$ as a comoving displacement vector and not as the
radius vector in proper coordinates.  Proper coordinates are not
used in this Appendix.

\subsection{Cross-covariance between Variables and Constraints}

Finally, we have the cross-covariances between our primary
variables and the constraints, i.e. the components of
$\left\langle Y_B\otimes Y_A \right\rangle$.  These follow from
Eqs.~(\ref{fourier})--(\ref{smoothdel}):
\begin{subequations}\label{covyab}
\begin{eqnarray}
  \left\langle\delta(\vec x\,)\Delta_0\right\rangle &=& \bar\xi(r)\equiv
    \int\frac{d^3k}{(2\pi)^3}\,P(k)W_R(k)j_0(kr)\ ,\quad r=\vert\vec x\,\vert\ ,
    \label{cov10}\\
  \left\langle\vec\psi(\vec x\,)\Delta_0\right\rangle&=&-\bar\eta(r)\vec x\ ,
    \quad\bar\eta(r)\equiv\int\frac{d^3k}{(2\pi)^3}\,P(k)W_R(k)
    \frac{j_1(kr)}{kr}\ ,\label{covi0}\\
  \left\langle\delta(\vec x\,)\vec\Delta_0\right\rangle&=&-\frac{d\bar\xi}{dr}
    \hat r\ ,\quad\hat r=\vec x/r\ ,\label{cov1j}\\
  \left\langle\vec\psi(\vec x\,)\otimes\vec\Delta_0\right\rangle&=&
    r\frac{d\bar\eta}{dr}\,\hat r\otimes\hat r+\bar\eta\,{\bf I}
    \label{cov1ij}\ .
\end{eqnarray}
\end{subequations}
To arrive at the last expression above, we have used the identity
\begin{equation}
  \label{nntens}
  \int\frac{d\Omega}{4\pi}\,e^{-i\vec k\cdot\vec x}\,\hat n\otimes
    \hat n=r\frac{\partial}{\partial r}\left[\frac{j_1(kr)}{kr}
    \right]\hat r\otimes\hat r+\frac{j_1(kr)}{kr}\,{\bf I}\ ,
\end{equation}
where $\vec k=k\hat n$, $\vec x=r\hat r$ and ${\bf I}$ is the
unit tensor.

\subsection{Mean of Variables Subject to Constraints}

Using Eqs.~(\ref{meany}) and (\ref{varcon})--(\ref{covyab}), we
get the mean values subject to the extremum constraints
$(\Delta_0,\vec\Delta_0=0)$,
\begin{subequations}\label{meanext}
\begin{eqnarray}
  \l \delta_1 \r_C & \equiv &
      \left\langle\delta_1 \vert \Delta_0,\vec\Delta_0 \right\rangle
    =\frac{\Delta_0}{\sigma_0^2}\bar\xi(r_1)\ ,\quad r_1=
      \vert\vec x_1\vert\ , \label{m1C}\\
  \l \delta_2 \r_C &\equiv &
     \left\langle\delta_2 \vert \Delta_0,\vec\Delta_0 \right\rangle
    = \frac{\Delta_0}{\sigma_0^2}\bar\xi(r_2)\ ,\quad r_2=
        \vert\vec x_2\vert\ , \label{m2C}\\
  \l \vec\psi_1 \r_C & \equiv &
     \left\langle\vec\psi_1 \vert \Delta_0,\vec\Delta_0 \right\rangle
    = -\frac{\Delta_0}{\sigma_0^2}\bar\eta(r_1) \vec x_1\ .
      \label{miC}
\end{eqnarray}
\end{subequations}
The subscript $C$ denotes ``subject to constraints.''

Eqs.~(\ref{meanext}) would be unchanged had we dropped the
extremum condition and imposed instead the single constraint
$\Delta(0)=\Delta_0$; the constraint $\vec\Delta_0=0$ has no effect
on the mean values.  This follows because $\langle\Delta_0\vec
\Delta_0\rangle=0$ so that the only way that $\vec\Delta_0$ can enter
the mean values is through terms proportional to
$\vec\Delta_0/\sigma_1^2$ which vanish when the extremum constraint
is imposed.

\subsection{Covariance of Variables Subject to Constraints}

We now compute the covariance matrix of our variables
$Y_B=(\delta_1, \delta_2,\vec\psi_1)$ subject to the extremum
constraints $Y_A=(\Delta_0,\vec\Delta_0)$.  The covariances follow
from Eqs.~(\ref{covy})--(\ref{covyab}). Recalling the definition
$C(Y_B, Y_B) \equiv\left\langle \Delta Y_B\otimes\Delta Y_B\vert
Y_A\right\rangle$, we obtain
\begin{subequations}\label{covfun}
\begin{eqnarray}
 C(\delta_1,\delta_2) & = & \xi(r)-\frac{\bar\xi_1\bar\xi_2}
    {\sigma_0^2}-\frac{1}{\sigma_1^2}\frac{d\bar\xi}{dr_1}
    \frac{d\bar\xi}{dr_2}\hat r_1\cdot\hat r_2\ ,\quad
    \vec r \equiv \vec x_1-\vec x_2 \ ,\ \hat r_1\equiv
    \frac{\vec x_1}{r_1}\ ,\ \hat r_2\equiv\frac{\vec x_2}{r_2}\ ,
    \label{C12}\\
 C(\delta_2,\vec\psi_1) & = & -\eta(r) \vec r
    +\frac{\bar\xi_2\bar\eta_1\vec x_1}{\sigma_0^2}+\frac{1}{\sigma_1^2}
    \frac{d\bar\xi}{dr_2}\left[\left(\bar\xi_1-3\bar\eta_1\right)
    \left(\hat r_1\cdot\hat r_2\right)\hat r_1+\bar\eta_1\,\hat r_2 \right]\ ,
    \label{C2i}\\
 C(\delta_1,\vec\psi_1) & = & \frac{\bar\xi_1\bar\eta_1\vec x_1}{\sigma_0^2}
    +\frac{1}{\sigma_1^2}\frac{d\bar\xi}{dr_1}(\bar\xi_1-2\bar\eta_1)
    \hat x_1\ ,\label{C1i}\\
  C(\vec\psi_1,\vec\psi_1) & = & c_r\hat r_1\otimes\hat r_1
    + c_t ({\bf I}- \hat r_1\otimes\hat r_1) \ ,\nonumber\\
  c_r(r_1) &\equiv& \sigma_\psi^2-\left(\frac{r_1\bar\eta_1}
    {\sigma_0}\right)^2 - \left(\frac{\bar\xi_1-2\bar\eta_1}{\sigma_1}
    \right)^2\ ,\label{Cpar}\\
  c_t(r_1)&\equiv&\sigma_\psi^2-\left(\frac{\bar\eta_1}{\sigma_1}
  \right)^2\ ,\label{Cper}
\end{eqnarray}
\end{subequations}
where $\bar\xi_i\equiv\xi(r_i)$ and
$\bar\eta_1\equiv\bar\eta(r_1)$, and we have used the relation
$r_1(d\bar\eta_1/dr_1)=\bar\xi_1-3\bar\eta_1$.  Note that
$C(\delta_i,\vec\psi_1)$ is a vector and
$C(\vec\psi_1,\vec\psi_1)$ is a second-rank tensor.  Later we will
need the tensor $C^{-1}(\vec\psi_1,\vec\psi_1)$, which is the
matrix inverse of $C(\vec\psi_1,\vec\psi_1)$.

The role of constraints is easy to see in Eqs.~(\ref{covfun}). The
constraint on $\Delta_0$ is responsible for the terms proportional
to $\sigma_0^{-2}$ while the constraint on $\vec\Delta_0$ gives rise
to the terms proportional to $\sigma_1^{-2}$.  Therefore, the
constraints are easily removed by setting $\sigma_1^{-2}\to0$ (to
eliminate the constraint on $\vec\Delta_0$) or $\sigma_0^{-2}$ (to
eliminate the constraint on $\Delta_0$). As expected, in the absence
of any constraints, the covariance matrix recovers the same
expressions as the unconstrained covariance in Eqs.~(\ref{covyb})
and (\ref{covybb}).

We will need one more covariance, $C(\vec\psi_1,\vec\psi_0)$
subject to the extremum constraints where
$\vec\psi_0\equiv\vec\psi(0)$. Using Eqs.~(\ref{covy}),
(\ref{covyab}), and (\ref{nntens}), we get
\begin{equation}
  \label{c0}
   C(\vec\psi_1,\vec\psi_0)=\frac{d\gamma}{dr_1}
    \,\hat r_1\otimes\hat r_1+\frac{\gamma}{r_1}\,\left({\bf I}-
    \hat r_1\otimes\hat r_1\,\right)\ ,\quad
  \gamma(r_1)\equiv\int\frac{d^3k}{(2\pi)^3}\,P(k)k^{-3}\,j_1(kr_1)\ .
\end{equation}

\subsection{Probability Distributions}

Now we are ready to calculate the probability distribution
functions and means of various quantities needed in
\S~\ref{sec:dfgauss}. Specifically, we need $p(\vec\psi_1)$, $\l
\delta_1|\vec\psi_1\r_C$, $\l \delta_2|\vec\psi_1\r_C$, $\l
\delta_2\r_C$, and $\l \delta_1\delta_2 |\vec\psi_1\r_C$ that
appear in the key Eqs.~(\ref{f1p1}) and (\ref{f2cp1}) for $f(\vec
w_1,t)$ and $f_{2c}(\vec w_1,\vec x_2,t)$.  Eq.~(\ref{m2C})
already gives $\l\delta_2\r_C$ but we need to calculate the other
quantities.

First, the probability density distribution for the displacement
$p(\vec\psi_1)$ (subject to the extremum constraints) is a
three-dimensional Gaussian whose mean and covariance are given
above, implying
\begin{equation}
  \label{fpsi1}
  p(\vec\psi_1)=(2\pi)^{-3/2}\left(c_r c_t^2\right)^{-1/2}
    \exp\left[-\frac{1}{2}\left(\vec\psi_1- \langle\vec\psi_1
     \rangle_C \right)\cdot  C^{-1}(\vpsia,\vpsia)
    \cdot\left(\vpsia-  \langle\vpsia \rangle_C \right)
    \right]\ .
\end{equation}
Note that the argument of the exponential is a scalar --- the dot
denotes a contraction (dot product).

The joint distribution of displacement and density (subject to the
constraints) is a little more complicated to work out since this
requires inverting the $4\times4$ covariance matrix of
$(\delta_2,\vec\psi_1)$. The distribution factors into conditional
and marginal distributions,
$p(\delta_2,\vec\psi_1)=p(\vec\psi_1)p(\delta_2\vert\vec\psi_1)$.
Straightforward algebra gives the conditional distribution,
\begin{equation}
  \label{f2cond}
  p(\delta_2\vert\vec\psi_1)=(2\pi Q)^{-1/2}\exp\left\{-\frac{1}{2Q}
  \left[\delta_2-\l \delta_2\r_C - C(\delta_2,\vpsia)\cdot
  C^{-1}(\vpsia,\vpsia) \cdot \left(\vec\psi_1- \l \vec\psi_1 \r_C
   \right)\right]^2\right\}\ ,
\end{equation}
where
\begin{equation}
  \label{vardel2}
  Q\equiv C(\delta_2,\delta_2)- C(\delta_2,\vpsia)\cdot
    C^{-1}(\vpsia,\vpsia) \cdot C(\vpsia,\delta_2) \ .
\end{equation}
Note that the second term in $Q$ (and the similar term in eq.
\ref{f2cond}) is a scalar. Writing it out with indices and using
the summation convention, $C(\delta_2,\delta_2)-Q=
C_i(\delta_2,\vpsia)C^{-1}_{ij}(\vpsia,\vpsia)C_j(\vpsia,\delta_2)$.

Eqs.~(\ref{f2cond}) and (\ref{vardel2}) could also have been
derived by noting that $\l Y_B | Y_A\r_C = \l \db | \vpsia \r_C -
\l \db \r_C$ for $Y_A=\vec\psi_1-\langle\vec\psi_1\rangle_C$ and
$Y_B=\delta_2-\l\delta_2\r_C$.  Then using Eq.~(\ref{meany})
and relations $\l Y_B Y_A\r _C=\l (\db - \l \db\r _C) (\vpsia - \l
\vpsia \r _C) \r_C = C(\db,\vpsia)$ and $\l Y_A Y_A\r _C=
C(\vpsia,\vpsia)$, we obtain
\begin{eqnarray}
  \label{d2psi1}
    \l \db | \vpsia \r _C &=& \l  \db \r _C + C(\db,\vpsia)\cdot
    C^{-1}(\vpsia,\vpsia) \cdot (\vpsia - \l \vpsia \r _C) \nonumber\\
      &=& \l\delta_2 \r_C -C(\delta_2,\vpsia)\cdot\frac{\partial}{\partial
       \vec\psi_1}\log p(\vec\psi_1)\ ,
\end{eqnarray}
which is identical to the mean in Eq.~(\ref{f2cond}).  The
constrained mean $\l\delta_1|\psi_1\r_C$ follows simply by
replacing $\l\delta_2\r_C$ with $\l\delta_1\r_C$ and $C(\delta_2,
\vec\psi_1)$ with $C(\delta_1,\vec\psi_1)$.

The only remaining constrained mean that we need to evaluate is
$\l \delta_1\delta_2 |\vec\psi_1\r_C$.  Direct evaluation of the
conditional distribution would require inverting a $5\times5$
matrix.  It is much simpler to use Eqs.~(\ref{meany}) and
(\ref{covy}) with $Y_B = \{(\da - \l \da \r _C),(\db - \l \db\r
_C)\}$ and $Y_A=\vpsia - \l \vpsia \r _C$.  From the off-diagonal
element of $\l\Delta Y_B\otimes\Delta Y_B\vert Y_A\r$, we get
\begin{equation}
  \label{d1d2psi1}
  \langle\delta_1\delta_2\vert\vec\psi_1\rangle _C=
    \l \delta_1 \r_C \l \delta_2 \r_C +
    C(\da,\db) - C(\da,\vpsia) \cdot
   C^{-1}(\vpsia,\vpsia)\cdot  C(\vpsia,\db) \ .
\end{equation}

At last we are ready to evaluate the one- and two-particle
distribution functions for small-amplitude Gaussian fluctuations.
Substituting $\l\delta_1|\vec\psi_1 \rangle_C$ into
Eq.~\ref{f1p1}), to second order in $\delta$ we obtain
\begin{equation}
  \label{f1lin}
  f(\vec w,t)=\bar\rho(1+\langle\delta\rangle_C)\,p(\vec\psi-
    C(\delta,\vec\psi\,)\,)\,(Hb)^{-3}+O(\delta^3)\ ,
\end{equation}
where $\l\delta\r_C$ is given by Eq.~(\ref{m1C}) with $r_1=r$ and
$C(\delta,\vec\psi\,)$ is a vector equal to the covariance of the
constrained $\delta$ and $\vec\psi$ given in Eq.~(\ref{C1i}).
Without an initial constraint we would have
$\langle\delta\rangle_C=0$ and $C(\delta,\vec\psi\,)=0$. An
initial constraint on the proto-halo shifts the mean density and
causes the density and velocity to be correlated hence it changes
the phase space density. Note that Eq.~(\ref{f1lin}) implies that
in linear theory the spatial and velocity distributions decouple.
As expected, the mean spatial density is
$\bar\rho(1+\langle\delta\rangle_C)$. The velocity distribution at
each point in the halo is simply the underlying Gaussian of
Eq.~(\ref{fpsi1}) with the mean shifted by $C(\delta,\vec\psi\,)$.
Note that the value of the proto-halo smoothed density constraint
$\Delta_0$ affects the density profile,
$\l\delta\r_C\propto\Delta_0$, but it does not affect the velocity
distribution up to second order in perturbation theory. The
imposition of the extremum constraint $\vec\Delta_0=0$ affects the
phase space density only through the term in
$C(\delta,\vec\psi\,)\propto\sigma_1^{-2}$.  Setting
$\sigma_1^{-2}\to0$ everywhere removes the constraint on
$\vec\Delta_0$.

The two-particle correlation is found by substituting
Eqs.~(\ref{d2psi1}) and (\ref{d1d2psi1}) into Eq.~(\ref{f2cp1}).
The result is
\begin{eqnarray}
  \label{f2cexact}
  f_{2c}(\vec w_1,\vec x_2,t) &=& \bar\rho^2\left\{
  C(\da,\db) - C(\da,\vpsia)\cdot C^{-1}(\vpsia,\vpsia)\cdot C(\vpsia,\db)
     \right.    \nonumber\\
  && \left. - \left[ C(\db,\vpsia) - C(\da,\vpsia)\l\delta_2\r_C\right]
   \cdot\frac{\partial}{\partial\vpsia} \log p(\vpsia) \right\}
    p(\vpsia)(Hb)^{-3}\ .
\end{eqnarray}
This equation is the main result of this Appendix.  It is exact
for Gaussian random fields.

\section{Stationary Solutions and Relaxation to Equilibrium}
\label{sec:stationary}

Stationary solutions of the Fokker-Planck equation are relevant
for systems that relax on a timescale faster than external changes
take place.  The stationary solution for classical Brownian motion
was given in \S~\ref{sec:brownian}.  Stationary solutions may
also exist for cosmological systems where the constant-temperature
heat bath is replaced by small-scale structure that rains upon
dark matter halos.

Stationary solutions have $\partial f/\partial t=0$ and $\vec F
\equiv \vec Af-{\bf D}\cdot\partial f/\partial\vec v=0$ so that
the flux vanishes in velocity space. We consider spherically
symmetric solutions as appropriate for the average halo.  From
Jeans' Theorem, the distribution function is a function of the
integrals of motion \citep{bt88}. For spherical systems it is
often assumed that stationary solutions have the form $f=f(E,L)$
where
\begin{equation}\label{eandl}
  E=\frac{1}{2}v^2+\Phi(\vec r\,)\ ,\ \
  L=\vert\vec r\times\vec v\,\vert\ .
\end{equation}
(The particle masses are irrelevant, so we use the specific energy and
angular momentum.) The collisionless Boltzmann equation gives no
constraint on $f(E,L)$.  The Fokker-Planck equation, however,
constrains $f$ because stationarity requires that ${\bf D}\cdot
\partial\ln f/\partial\vec v=\vec A$.  Assuming spherical symmetry, we
write
\begin{equation}\label{driftvec}
  \vec A=\alpha\hat r-\gamma\vec v\ ,
\end{equation}
introducing the radial drift coefficient $\alpha$ and drag
coefficient $\gamma$. We assume isotropic diffusivity $D$.  For
the globular cluster case with Chandrasekhar's dynamical friction,
$\alpha=0$.  For the case of quasilinear cosmological fluctuations
arising from Gaussian random fields, we have seen that $\gamma=0$.
(The cosmological case also allows for different diffusivities in
the radial and tangential directions, but that is an unnecessary
complication here.)

Requiring the phase space flux to vanish now gives the following
constraints on the distribution function $f(E,L)$:
\begin{equation}\label{station0}
  \frac{\partial\ln f}{\partial E}=-\frac{\gamma}{D}
    +\frac{\alpha}{v_rD}\ ,\ \
  \frac{r^2}{L}\frac{\partial\ln f}{\partial L}=-\frac{\alpha}{v_r}{D}\ .
\end{equation}
It is interesting to see what conditions lead to thermal
equilibrium with a Maxwellian distribution of velocities, i.e.
Eq.~(\ref{equilsol}) with $f=W$.  Eq.~(\ref{station0}) shows that
the conditions are $\alpha=0$ and $\gamma/D=m/(k_{\rm B}T)$. Thus,
the drift must be velocity dependent and behave like a drag, $\vec
A\propto-\vec v$. Friction converts shear into heat in such a way
as to drive the system towards thermal equilibrium. Chandrasekhar
dynamical friction satisfies these conditions; therefore, two-body
relaxation drives the velocity distribution towards the
Maxwell-Boltzmann form.  Quasilinear cosmological fluctuations,
however, are quite different, having $\gamma=0$ and $\alpha\ne0$,
implying $\partial\ln f/\partial E=-(r^2/L)
\partial\ln f/\partial L=\alpha/(v_rD)$.  While these equations
do not have simple general solutions, it is clear that
cosmological fluctuations do not drive a system towards thermal
equilibrium with equipartition of energies.  Our cosmological
relaxation process is similar to violent relaxation in this
respect \citep{dlb67}.

Although the general cosmological case of relaxation in a dark
matter halo with $\alpha\ne0$ is complicated and cannot be fully
solved here, there is a special case worthy of closer examination,
namely, the evolution of an unconstrained cosmological Gaussian
random field, where fluctuations generate peculiar velocities.
That is, we drop the constraint of having a halo centered at $r=0$
and instead consider the velocity distribution of dark matter
particles at a randomly selected point in space. To consider the
cosmological case, we should use the appropriate comoving
variables to factor out the Hubble expansion. In place of $\vec r$
and $\vec v$, we use the following variables:
\begin{equation}\label{comovar}
  \vec x=a^{-1}\vec r\ ,\ \ \vec u=a(\vec v-H\vec r\,)\ ,
\end{equation}
where $a(t)$ is the cosmic scale factor and $H=d\ln a/dt$.  With
this change of variables, the Fokker-Planck equation (\ref{fplin})
becomes
\begin{equation}\label{fpcomov}
  \frac{\partial f}{\partial t}+\frac{\vec u}{a^2}\cdot\frac
    {\partial f}{\partial x}+a\left(\vec g-\frac{1}{a}\frac{d^2a}
    {dt^2}\vec r\,\right)\cdot\frac{\partial f}{\partial\vec u}
  =a\frac{\partial}{\partial\vec u}\cdot\left(-\vec A f+a{\bf D}
    \cdot\frac{\partial f}{\partial\vec u}\right)\ .
\end{equation}
If the velocities are measured in the comoving frame, $\vec g_{\rm
T}$ is replaced by $\vec g$.  When using comoving coordinates the
gravitational field is reduced by the Hubble acceleration,
$(d^2a/dt^2)\vec x$.  If the mean density field is homogeneous,
then all particles move with the unperturbed Hubble expansion,
$\vec r\propto a(t)$, implying $\vec g=(d^2a/dt^2)\vec x$. Thus,
both the velocity and acceleration terms on the left-hand side
vanish when $\partial f/\partial\vec x=0$. If we apply no initial
constraint on the density field, then the one-point distribution
function $f$ is a function only of time and peculiar velocity
$\vec v-H\vec r$. We use the variable $\vec u$ for the velocity
because $\vec v-H\vec r$ decreases in proportion to $a^{-1}(t)$
for a freely-falling body in an unperturbed homogeneous and
isotropic expanding universe.  If not for the right-hand side, the
solution of Eq.~(\ref{fpcomov}) for a homogeneous and isotropic
model would be $f(\vec x, \vec u,t)=f(\vec x=\vec 0,\vec u,t=0)$,
i.e. the the initial peculiar velocity distribution would simply
redshift with cosmic expansion and remain spatially homogeneous.

The Fokker-Planck equation does not, however, describe an
unperturbed homogeneous and isotropic universe.  Rather it
describes the statistical average of an ensemble of universes with
fluctuations.  In the case of unconstrained initial conditions,
$f=f(u,t)$ and Eq.~(\ref{fpcomov}) reduces to
\begin{equation}\label{fpcomhom}
  \frac{1}{a}\frac{\partial f}{\partial t}=\frac{1}{u^2}
    \frac{\partial}{\partial u}\left[u^2\left(\alpha_u f+aD
    \frac{\partial f}{\partial u}\right)\right]\ ,
\end{equation}
where $\alpha_u=\alpha_u(u,t)$ and $D=D(u,t)$ in general.  We have
used statistical isotropy to require $\vec A=-\alpha_u\hat u$
where $\hat u$ is a unit vector in the direction of $\vec u$. In
second order perturbation theory (\S~\ref{sec:fpdiffco}),
$\alpha_u=0$ (no drag) and $D=4\pi G\bar\rho aHb\sigma_\psi^2$
(assuming that we measure velocities in the comoving frame.) The
Fokker-Planck equation then takes the form of a diffusion equation
with $D$ independent of $u$, and the general solution is
\begin{equation}\label{diffsol}
  f(\vec u,t)=\int\exp\left(-\frac{\vert\vec u-\vec u^{\,\prime}
    \vert^2}{4\int_0^ta^2D\,dt}\right)\,\frac{f(\vec u,0)\,d^3u'}
    {\left(4\pi\int_0^ta^2D\,dt\right)^{3/2}}\ .
\end{equation}
The effect of fluctuations is simply to spread the initial
velocity distribution (if $D>0$) while retaining the Gaussian
form. Physically, the gravitational fluctuations of substructure
cause local heating.

Contrary to Eq.~(\ref{diffsol}), cosmological simulations show
that the distribution of peculiar velocities in the nonlinear
regime is closer to exponential rather than Gaussian, at least in
the tails \citep{sd01}. It is interesting to ask whether the
Fokker-Planck equation can explain such a result.  With the drift
term retained, Eq.~(\ref{fpcomhom}) has the stationary solution
\begin{equation}\label{fpstat}
  f(\vec u,t)=\exp-\int^u\frac{\alpha_u(u,t)\,du}{a(t)D(u,t)}\ .
\end{equation}
If $\alpha_u/D$ is positive and independent of $u$, the result is
indeed an exponential distribution of peculiar velocities $u$.  In
second-order perturbation theory we found that $\alpha_u=0$ and
$D$ is independent of $u$, however Chandrasekhar's dynamical
friction gives $A/D\propto u$.  Evidently, a third case is needed
to explain the exponential velocity distribution --- there must be
a drag such that $\alpha_u/D$ is independent of $u$, quite unlike
dynamical friction from two-body relaxation. Determining whether
such a drag term arises from nonlinear gravitational clustering is
a subject for further work.

As one final example, we consider the effects of both drag and
radial drift in a simplified model illustrating the effects of
collisions on a system for which the relaxation time is much
longer than the dynamical time (e.g. the crossing-time for a
particle in a halo).  Under these conditions, the system at all
times is nearly in collisionless equilibrium with the phase space
density being a function of the actions alone (and not the angles,
using action-angle variables).  Because the actions are conserved
in the absence of collisions, for quasi-static evolution
Eq.~(\ref{fplin}) simplifies to
\begin{equation}\label{fpqs}
  \frac{\partial f}{\partial t}=-\frac{\partial}{\partial\vec v}
    \cdot\left[(\vec\alpha-\gamma\vec v\,)f-{\bf D}\cdot\frac
    {\partial f}{\partial\vec v}\right]\ ,
\end{equation}
where we have generalized the radial drift to an arbitrary vector
field $\vec\alpha$ which we assume to be independent of $\vec v$.
We can solve this equation exactly for the simple case where
$\vec\alpha$, $\gamma$, and ${\bf D}$ are all constants.  The
general solution is the Green's function solution
\begin{equation}\label{fpqssol}
  f(\vec r,\vec v,t)=\int d^3v'\,G(\vec r,\vec v-\vec v^{\,\prime},t)
    \,f(\vec r,\vec v^{\,\prime},0)\ ,
\end{equation}
with Green's function
\begin{equation}\label{fpqsgreen}
  G(\vec r,\vec v,t)=\int\frac{d^3s}{(2\pi)^3}\,\exp\left\{i\vec s
    \cdot\left[\vec v-\frac{1}{\gamma}\left(1-e^{-\gamma t}\right)\vec
    \alpha\right]-\frac{1}{2\gamma}\vec s\cdot{\bf D}\cdot\vec s
    \left(1-e^{-2\gamma t}\right)\right\}\ .
\end{equation}
The parameter $\gamma^{-1}$ is a relaxation time.
Eq.~(\ref{fpqsgreen}) simplifies to a normalized Gaussian in two
limits,
\begin{equation}\label{fpqsgrlim}
  G(\vec r,\vec v,t)=\begin{cases}
    {\cal N}(\vec\alpha t,2{\bf D}t)\ ,&
    \gamma t\ll1 ,\cr {\cal N}(\vec\alpha/\gamma,{\bf D}/\gamma)
    \ ,& \gamma t\gg1\ ,\cr
  \end{cases}
\end{equation}
where ${\cal N}(\vec u,{\bf M})$ is a multivariate normal with
mean $\vec u$ and covariance matrix ${\bf M}$.  The solution for
$\vec\alpha=\gamma=0$ was given previously in Eq.~(\ref{diffsol}).
Now we have a more general solution showing that the equilibrium
(for constant diffusion coefficients) is a Maxwellian with
velocity dispersion tensor ${\bf D}/\gamma$ just in
Eq.~(\ref{fpstat}) if $\alpha_u=\gamma u$, but now the drift
vector $\vec\alpha$ produces an offset of the mean velocity
$\langle\vec v\rangle= \alpha/\gamma$. Thus, velocity-independent
drift $\alpha$ and velocity-dependent drag $-\gamma\vec v$ play
completely different roles in collisional relaxation.  In later
papers we will investigate these roles for more realistic models
of nonlinear halo evolution.

\end{document}